\documentclass[useAMS,usenatbib]{mn2e}
\usepackage{graphicx,url,multirow,mathrsfs,amssymb,amsmath,relsize}
\def\apj{ApJ}%
\def\apjl{ApJ}%
\def\aap{A\&A}%
\def\aapr{A\&A~Rev.}%
\def\mnras{MNRAS}%
\def\nat{Nature}%
\newcommand{\eqb}{\begin{eqnarray}}
\newcommand{\eqe}{\end{eqnarray}}
\newcommand{\std}{\operatorname{std}}
\def\rg{$r_{\rm g}$}
\def\ms{M$_{\rm \sun}$}
\newcommand{\xmm}{{\it XMM--Newton }}
\newcommand{\xmmp}{{\it XMM--Newton}}
\title[X-ray reprocessing echoes in the PSDs of AGN]{A search for X-ray reprocessing echoes in the power spectral density functions of AGN\thanks{Based on observations obtained with \xmmp, an ESA science mission with instruments and contributions directly funded by ESA Member States and NASA.}}

\author[D.~Emmanoulopoulos et al.]{D.~Emmanoulopoulos,$^{1,2}$\thanks{E-mail: D.Emmanoulopoulos@soton.ac.uk} I.~E.~Papadakis,$^{2,\,3}$ A.~Epitropakis,$^{2}$  T.~Pech\'a\v{c}ek,$^{2,\,4}$ \newauthor M. Dov\v{c}iak,${^4}$ and I.~M.~M\textsuperscript{c}Hardy$^{1}$\\
$^{1}$Physics and Astronomy, University of Southampton, Southampton, SO17 1BJ, UK\\
$^{2}$Physics Department and Institute of Theoretical and Computational Physics, University of Crete, PO Box 2208, 71003 Heraklion, Greece\\
$^{3}$IESL, Foundation for Research and Technology-Hellas, 71110 Heraklion, Crete, Greece\\
$^{4}$Astronomical Institute, Academy of Sciences, Bo\v{c}n\'{\i}~II 1401, CZ-14131 Prague, Czech Republic}

\date{Accepted 2016 June 2. Received 2016 June 2; in original form 2016 February 29}
\pagerange{\pageref{firstpage}--\pageref{lastpage}}\pubyear{2015}

\begin{document}
\label{firstpage}
\maketitle

\begin{abstract}
We present the results of a detailed study of the X-ray power spectra density (PSD) functions of twelve X-ray bright AGN, using almost all the archival \xmm data. The total net exposure of the EPIC-pn light curves is larger than 350 ks in all cases (and exceeds 1 Ms in the case of 1H\;0707-497). In a physical scenario in which X-ray reflection occurs in the inner part of the accretion disc of AGN, the X-ray reflection component should be a filtered echo of the X-ray continuum signal and should be equal to the convolution of the primary emission with the response function of the disc. Our primary objective is to search for these reflection features in the $5-7$ keV (iron line) and $0.5-1$ keV (soft) bands, where the X-ray reflection fraction is expected to be dominant. We fit to the observed periodograms two models: a simple bending power law model (BPL) and a BPL model convolved with the transfer function of the accretion disc assuming the lamp-post geometry and X-ray reflection from a homogeneous disc. We 
do not find any significant features in the best-fitting BPL model residuals either in individual PSDs in the iron band, soft and full band ($0.3-10$ keV) or in the average PSD residuals of the brightest and more variable sources (with similar black hole mass estimates). The typical amplitude of the soft and full-band residuals is around $3-5$ per cent.  It is possible that the expected general relativistic effects are not detected because they are intrinsically lower than the uncertainty of the current PSDs, even in the strong relativistic case in which X-ray reflection occurs on a disc around a fast rotating black hole having an X-ray source very close above it. However, we could place strong constrains to the X-ray reflection geometry with the current data sets if we knew in advance the {\it intrinsic} shape of the X-ray PSDs, particularly its high frequency slope.
\end{abstract}


%%%%%%%%%%%%%%%%%%%%%%%%%%%%%%%%%%%%%%%%%%%%%%%%%%

%%%%%%%%%%%%%%%%% BODY OF PAPER %%%%%%%%%%%%%%%%%%

\section{INTRODUCTION}
\label{ref:intro}

\begin{table*}
\label{tab:obj}
\begin{minipage}{170mm}
\caption{The sample and the \xmm observations. Columns (1) list the source name and BH mass estimate (together with the corresponding reference). The number in square brackets, next to the AGN name, indicates the number of the 10 ks light curve segments that we used for the PSD estimation. Columns (2) list the \xmm observation IDs, and columns (3) list the net exposure of the observations that we used (i.e\ the total light curve duration after background subtraction and screening).} \vspace{0em}
\label{tab:obs}
\hspace*{1em}\begin{tabular}{@{}cccc}
\hline
(1) & (2) & (3) \\
\underline{AGN name} & Obs ID & Net exp. \\
 BH Mass  ($\times 10^6$ \ms)    &    &       (ks)    \\
\hline

\underline{NGC\;7314} [36]\\
$0.87\pm0.45$ 	&	0111790101	&	43.3	\\
\citet{mchardy13} &	0311190101	&	77.5	\\
&	0725200101	&	124.7	\\
&	0725200301	&	130.6	\\

\underline{NGC\;4051} [51] \\
$1.73^{+0.55}_{-0.52}$  &	0109141401	&	103.0	\\
\citet{denney10} &	0157560101	&	50.0	\\
&	0606320101	&	45.3	\\
&	0606320201	&	42.0	\\
&	0606320301	&	21.1	\\
&	0606320401	&	18.9	\\
&	0606321301	&	30.2	\\
&	0606321501	&	34.0	\\
&	0606321601	&	41.5	\\
&	0606321701	&	38.4	\\
&	0606321801	&	18.8	\\
&	0606322001	&	22.1	\\
&	0606322101	&	29.2	\\
&	0606322201	&	30.8	\\
&	0606322301	&	42.3	\\
\underline{Mrk\;766} [55]	\\
$1.76^{+1.56}_{-1.40}$  &	0109141301	&	116.9	\\
\citet{bentz09} &	0304030101	&	95.1	\\
&	0304030301	&	98.5	\\
&	0304030401	&	93.0	\\
&	0304030501	&	74.7	\\
&	0304030601	&	85.2	\\
&	0304030701	&	29.1	\\

\underline{Ark\;564} [50]        \\
$2.32\pm0.41$\footnote{From equation 5 of \citet{vestergaard06} with mean values of ${\rm FWHM(H\,_\beta)}$ and $\lambda{\rm L}_\lambda(5100$ \AA$)$ from \citet{romano04}.}     &	0206400101	&	98.7	\\
&	0670130201	&	59.1	\\
&	0670130301	&	55.5	\\
&	0670130401	&	56.7	\\
&	0670130501	&	66.9	\\
&	0670130601	&	57.0	\\
&	0670130701	&	43.5	\\
&	0670130801	&	57.8	\\
&	0670130901	&	55.5	\\
&	0006810101	&	10.6	\\
\underline{1H\;0707-495} [121]    \\
$2.34\pm0.71$   &	0110890201	&	40.7	\\
\citet{zhou05} &	0148010301	&	78.1	\\
&	0506200201	&	38.7	\\
&	0506200301	&	38.7	\\
&	0506200401	&	40.6	\\
&	0506200501	&	40.9	\\
&	0511580101	&	121.6	\\
&	0511580201	&	102.1	\\
&	0511580301	&	104.1	\\
&	0511580401	&	101.8	\\
&	0554710801	&	96.1	\\
&	0653510301	&	113.8	\\
&	0653510401	&	125.7	\\
&	0653510501	&	116.9	\\
&	0653510601	&	119.5	\\

\end{tabular}
\hspace*{8em}\parbox{1\linewidth}{\vspace*{-8em}\par
\begin{tabular}{@{}cccc}
\hline
(1) & (2) & (3) \\
\underline{AGN Name} & Obs ID   & Net exp. \\
BH Mass  ($\times 10^6$ \ms) &                      &  (ks)    \\
\hline

\underline{MCG--6-30-15} [68]   \\
$4.5^{+1.5}_{-1.0}$ &	0693781201	&	127.2	\\
\citet{mchardy13} &	0693781301	&	129.7	\\
&	0693781401	&	48.5	\\
&	0111570101	&	33.1	\\
&	0111570201	&	53.0	\\
&	0029740101	&	80.6	\\
&	0029740701	&	122.5	\\
&	0029740801	&	124.1	\\

\underline{IRAS\;13224-3809} [34]  \\
$5.75\pm0.82$   &	0110890101	&	37.4	\\
\citet{zhou05}  &	0673580101	&	50.3	\\
&	0673580201	&	84.1	\\
&	0673580301	&	73.9	\\
&	0673580401	&	113.5	\\

\underline{SWIFT\;J2127+5654} [43]\\
$15$            &	0655450101	&	117.1	\\
\citet{malizia08} &	0693781701	&	134.6	\\
&	0693781801	&	129.7	\\
&	0693781901	&	71.4	\\

\underline{Mrk\;335} [35]     \\   
$26\pm8$        &	0101040101	&	31.6	\\
\citet{grier12} &	0306870101	&	122.5	\\
&	0510010701	&	16.8	\\
&	0600540501	&	80.7	\\
&	0600540601	&	112.3	\\

\underline{NGC\;3516} [40] \\
$31.7^{+2.8}_{-4.2}$    &	0107460601	&	89.4	\\
\citet{denney10} &	0107460701	&	122.2	\\
&	0401210401	&	51.8	\\
&	0401210501	&	60.8	\\
&	0401210601	&	61.7	\\
&	0401211001	&	36.2	\\

\underline{PG\;1211+143\footnote{Two observations (obs IDS: 0208020101 and 0502050201) were not used due to increase background activity.}} [62]\\
$146\pm44$ 	&	0112610101	&	53.1	\\
\citet{peterson04} &	0502050101	&	46.3	\\
&	0745110101	&	77.4	\\
&	0745110201	&	98.2	\\
&	0745110301	&	51.8	\\
&	0745110401	&	92.0	\\
&	0745110501	&	54.6	\\
&	0745110601	&	91.9	\\
&	0745110701	&	95.6	\\

\underline{PKS\;0558-504} [56] \\
$250^{+50}_{-190}$ 	&	0555170201	&	102.7	\\
\citet{gliozzi10} &	0555170301	&	111.3	\\
&	0555170401	&	123.3	\\
&	0555170501	&	123.1	\\
&	0555170601	&	111.0	\\ 
\end{tabular}}
\noindent\rule{17.3cm}{0.4pt}
\vspace*{-3em}
\end{minipage}
\end{table*}

It has been over 20 years since the first detection of iron emission lines in the X-ray spectra of active galactic nuclei (AGN), and the first observational evidence of their asymmetrical shape \citep[e.g.][]{nandra94,tanaka95}. Since then, the general and special relativity effects to the line's shape have been studied extensively \citep[e.g.][]{fabian89,laor91,dovciak04,brenneman06,dauser10}. The modelling of the iron line profiles, and of other X-ray reflection spectral features (like e.g.\ the `soft-excess' and the `Compton-hump'), have been an important tool for the investigation of the X-ray source/accretion disc geometry in X-ray bright AGN. Recently, the discovery of the so-called `X-ray reverberation time-lags' in many AGN e.g.\ \citet{fabian09,emmanoulopoulos11b,demarco13,zoghbi13,kara15} \citep[also see for a recent review][]{uttley14} has provided an extra support for the case of the X-ray illumination of the innermost regions of the accretion disc. The observed time delays suggest a small sized 
X-ray source located atop a disc, extending down to the innermost stable orbital radius (ISCO) around a supermassive black hole (BH), which is in many cases fast rotating \citep[e.g.][]{wilkins13,emmanoulopoulos14,cackett14,chainakun15}.

Recently, \citet{papadakis16} (P16 hereafter) proposed a new method to investigate the X-ray reprocessing scenario, based on the study of the X-ray power spectral density (PSD) functions. If the innermost parts of AGN are illuminated by the X-ray source, the observed X-rays should be the sum of the X-ray photons emitted by the source and those which are reprocessed by the disc. The latter component is a filtered echo of the X-ray continuum signal, and should be equal to the convolution of the primary emission with the response function of the disc. As a result, the observed power spectra should display features of this echo. 

P16 assumed the so-called lamp-post geometry, a neutral disc with solar iron abundance, and considered various X-ray source heights, inclinations, and spin values of the central BH. They found that the ratio between the observed PSDs over the intrinsic ones should show a prominent dip, followed by oscillations around a constant power level, with decreasing amplitude at higher frequencies. The frequency of the main dip is determined by both the BH mass and the X-ray source height, and it is energy independent. The amplitude of the dip increases with increasing BH spin and inclination angle, as long as the height of the lamp is smaller than 10 gravitational radii, \rg.

We present the results from a detailed X-ray PSD study of twelve AGN, observed by \xmmp. Our primary aim is to search for X-ray reverberation signals in the observed PSDs by fitting them with the theoretical models developed by P16. The reflection fraction is expected to be quite strong in the $5-7$ keV band, which includes most of the iron line emission at around 6.4 keV (the iron-line band hereafter), and the P16 results are directly applicable to the PSDs in this band. However, the signal-to-noise ratio of the light curves in the iron-line band is significantly smaller than the signal-to-noise ratio of light curves at lower energies of $0.5-1.5$ keV. Many AGN show emission in excess of the extrapolation of the power-law spectral component at such low energies, which can be more than twice the power-law continuum flux. This so-called `soft-excess' component is consistent with X-ray reflection from the inner parts of a mildly ionized accretion disc \citep[e.g.][]{crummy06}. We therefore decided to study PSD 
in both the iron-line and the $0.5-1$ keV band (the soft band hereafter), in order to search for relativistically induced echo features. 

In Sect.~\ref{sect:obs_datRed} we present our sample, the observations and the data reduction procedure for the construction of the light curves. Following that, in Sect.~\ref{sect:PSDestim} we outline the power spectrum estimation method together with our objective. In Sect.~\ref{sect:PSD_modelFit} we present in detail our model fitting approach to the observed PSDs, giving also the best-fitting results. In Sect.~\ref{sect:modelCompari} we compare the best-fitting results from the different models. Finally, in Sect.~\ref{sect:disc} we summarize our results and discuss possible implications. 

%%%%%%%%%%%%%%%%%%%%%%%%%%%%%%%%%%%%%%%%%
\section{OBSERVATIONS AND DATA REDUCTION}
%%%%%%%%%%%%%%%%%%%%%%%%%%%%%%%%%%%%%%%%%
\label{sect:obs_datRed}
Table~\ref{tab:obj} lists the objects that we study, together with the details about the corresponding \xmm observations. The first, second and third columns list the source name together with the BH mass estimate (M$_{\rm BH}$), the observation identifier (obs ID), and the net exposure of each observation, respectively. Note, that the objects are listed in order of increasing M$_{\rm BH}$. The number in the parenthesis, next to the source name, indicates the number of 10 ks segments that we use to estimate the PSDs (see Sect.~\ref{sect:PSDestim}).

The objects in our sample are X-ray bright AGN and have been extensively studied in X-rays and they were selected based on the large number of \xmm observations obtained for each one of them. We have used almost all observations in the \xmm archive for each object, except from few observations having a net exposure time less than 10 ks, as this is the minimum duration for the periodogram estimation (see Sect.~\ref{sect:PSDestim}). The total net exposure of the light curves is larger than 350 ks for all objects in the sample.

We processed the data from the \xmm satellite using the scientific analysis system \citep[{\sc sas} v. 14.0.0;][]{gabriel04}. We considered only EPIC-pn data \citep{struder01}. Source and background light curves were extracted from circular regions on the CCD, with the former having a fixed radius of 800 pixels (40 arcsec) centred at the source coordinates listed on the NASA/IPAC Extragalactic Database. The positions and radii of the background regions were determined by placing them sufficiently far from the location of the source, while remaining within the boundaries of the same CCD chip.

The source and background light curves were extracted in the $0.5-1$ and $5-7$ keV energy bands, with a bin size of $\Delta t=$100 s, using the {\sc sas} command \textit{evselect}. We included the criteria PATTERN$=0-4$ and FLAG$=$0 in the extraction process, which select only single and double pixel events, rejecting at the same time bad pixels from the edges of the detector's CCD chips. Periods of high background flaring activity were determined by inspecting the $10-12$ keV light curves (which should include a negligible number of source photons) and they were extracted from the whole surface of the detector. These periods were then excluded during the source and background light curve extraction process.

We checked all source light curves for pile-up using the {\sc sas} task \textit{epatplot}, and found that only observations 670130201, 670130501 and 670130901 of Ark\;564 are affected. For those observations we used annular instead of circular source regions with inner radii of 280, 200 and 250 pixels (the outer radii were held at 800 pixels), respectively, which we found to adequately reduce the effects of `pile-up'.

The background light curves were then subtracted from the corresponding source light curves using the {\sc sas} tool \textit{epiclccorr}. Most of the resulting light curves were continuously sampled, except for a few cases which contained a small number of missing observations. These were either randomly distributed throughout the duration of an observation, or appeared in groups of less than 10 points. We replaced the missing observations by linear interpolation, with the addition of the appropriate Poisson noise.

%%%%%%%%%%%%%%%%%%%%%%%%%%%%%%%%%%%%%%%%%
\section{POWER SPECTRAL DENSITY ESTIMATION}
%%%%%%%%%%%%%%%%%%%%%%%%%%%%%%%%%%%%%%%%%
\label{sect:PSDestim}

%%%%%%%%%% FIGURE 1
\begin{figure*}
\includegraphics[width=3.3in]{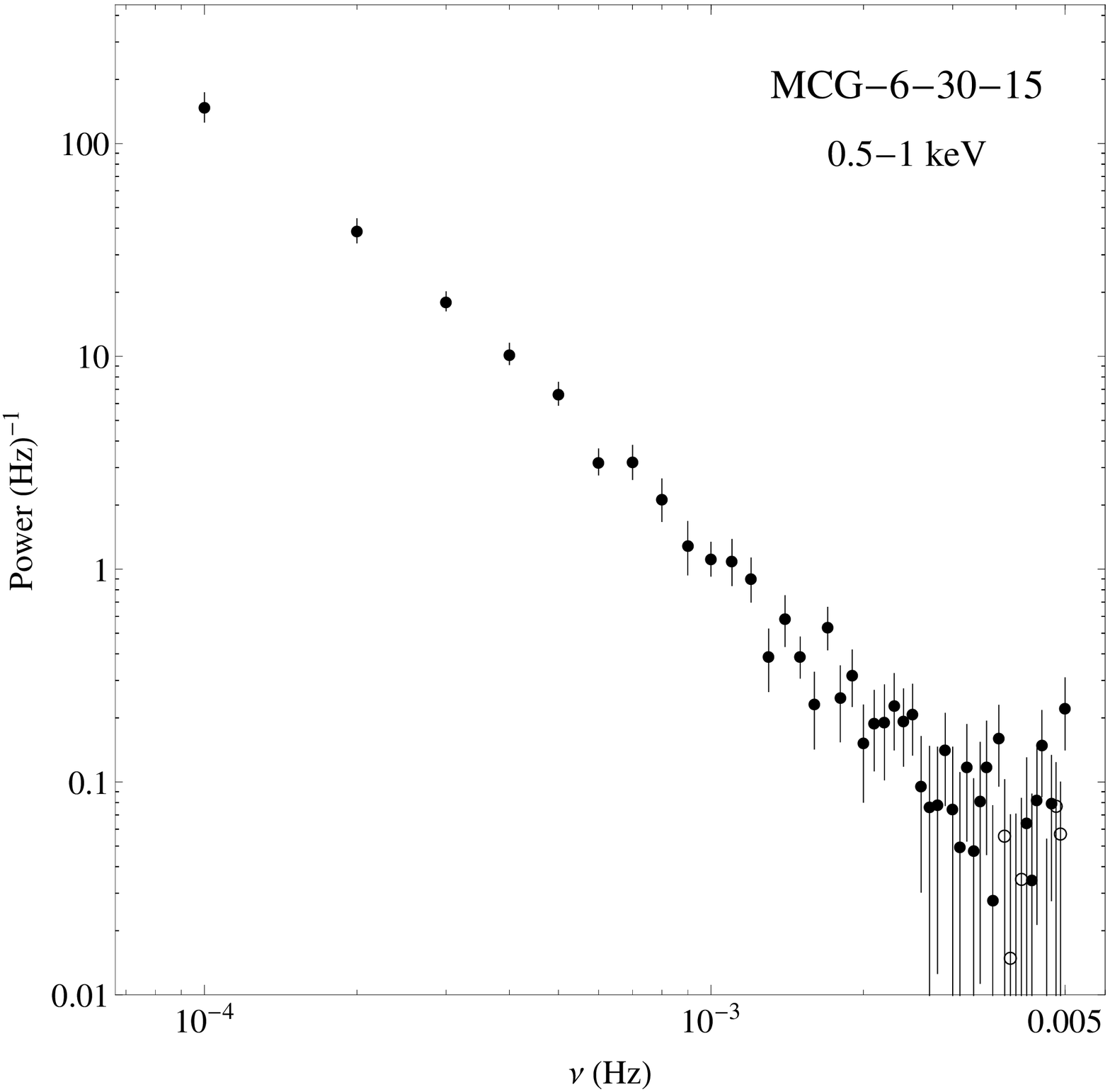}\hspace{1em}
\includegraphics[width=3.3in]{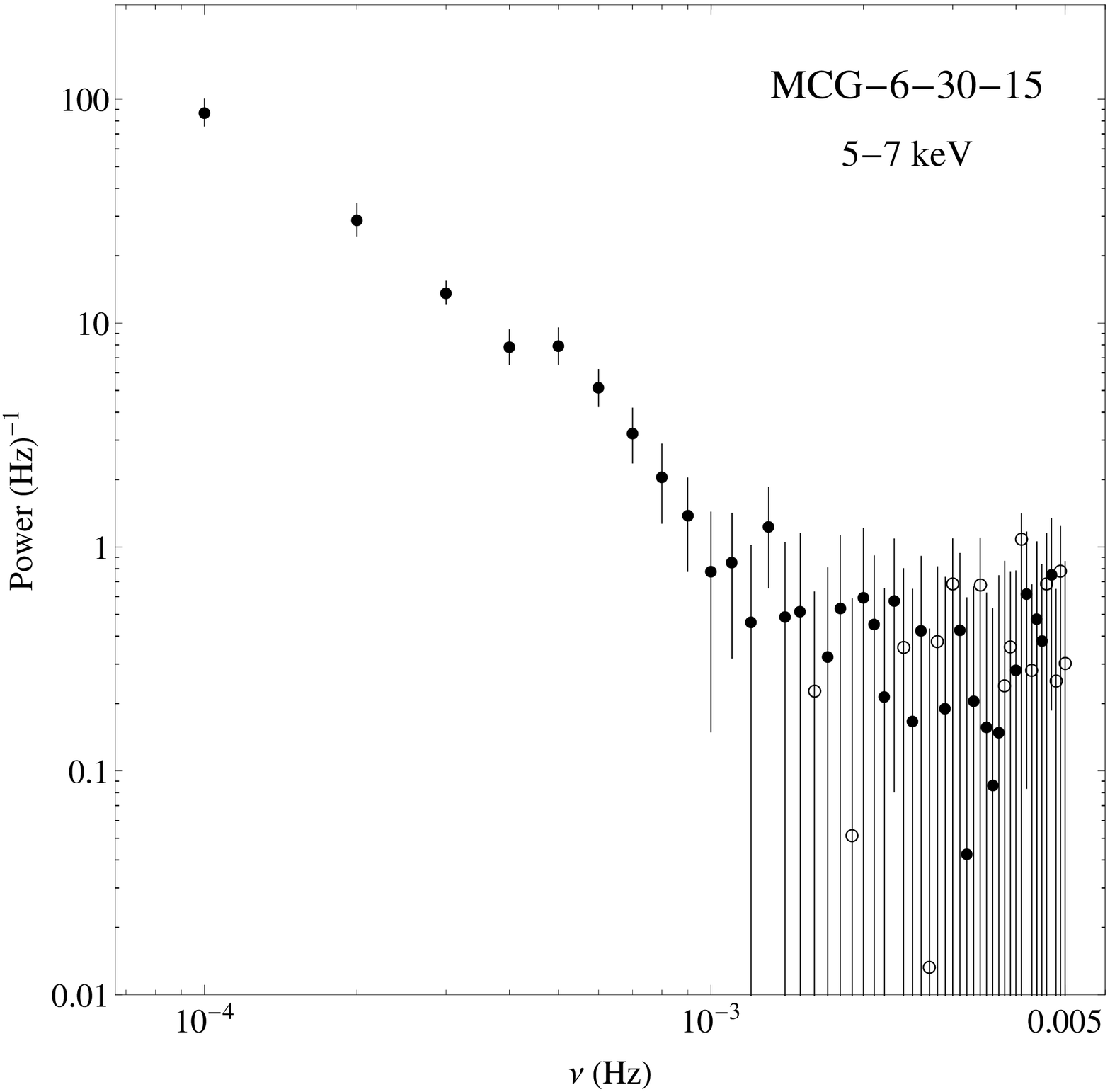}
\includegraphics[width=3.3in]{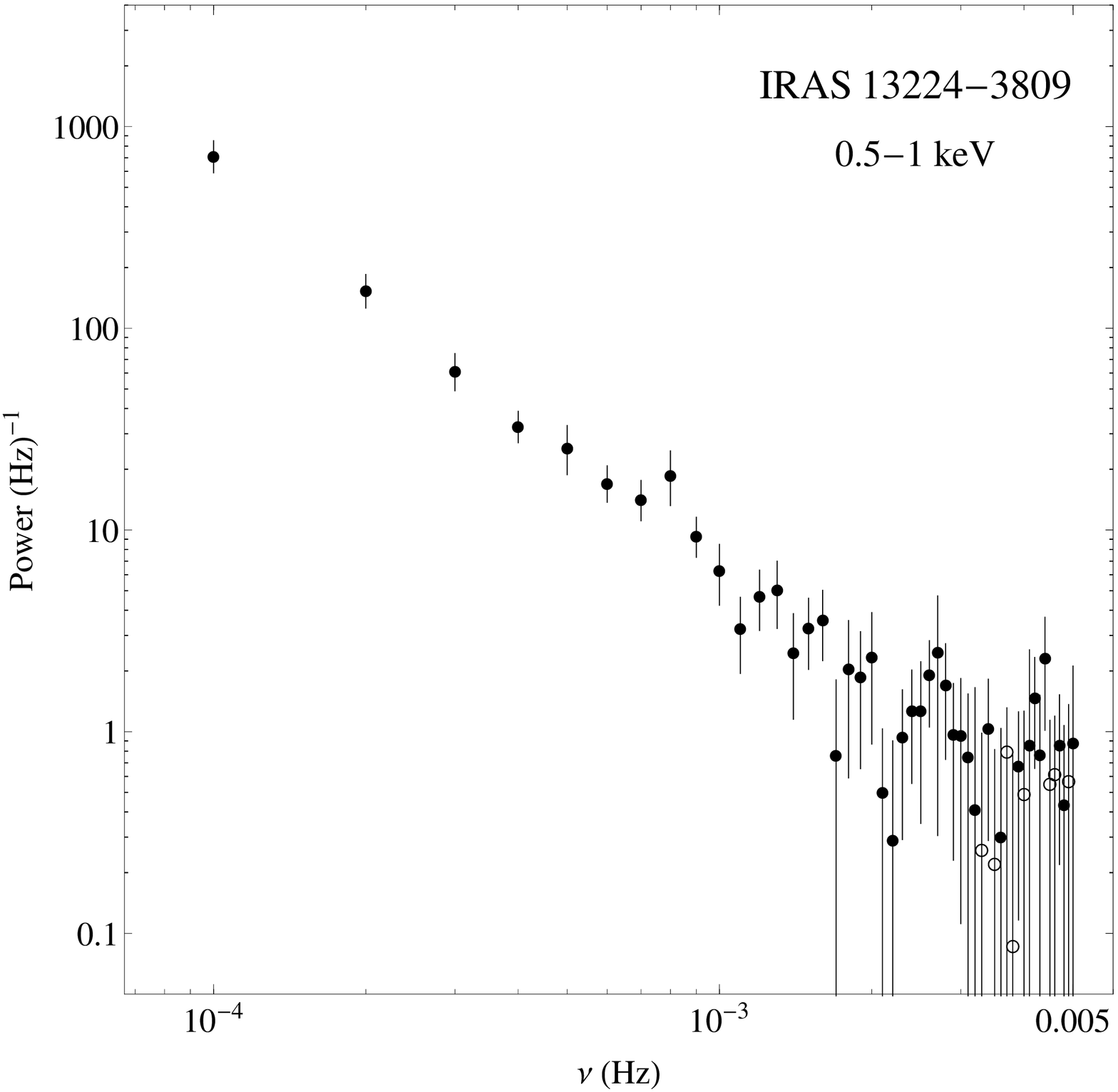}\hspace{1em}
\includegraphics[width=3.3in]{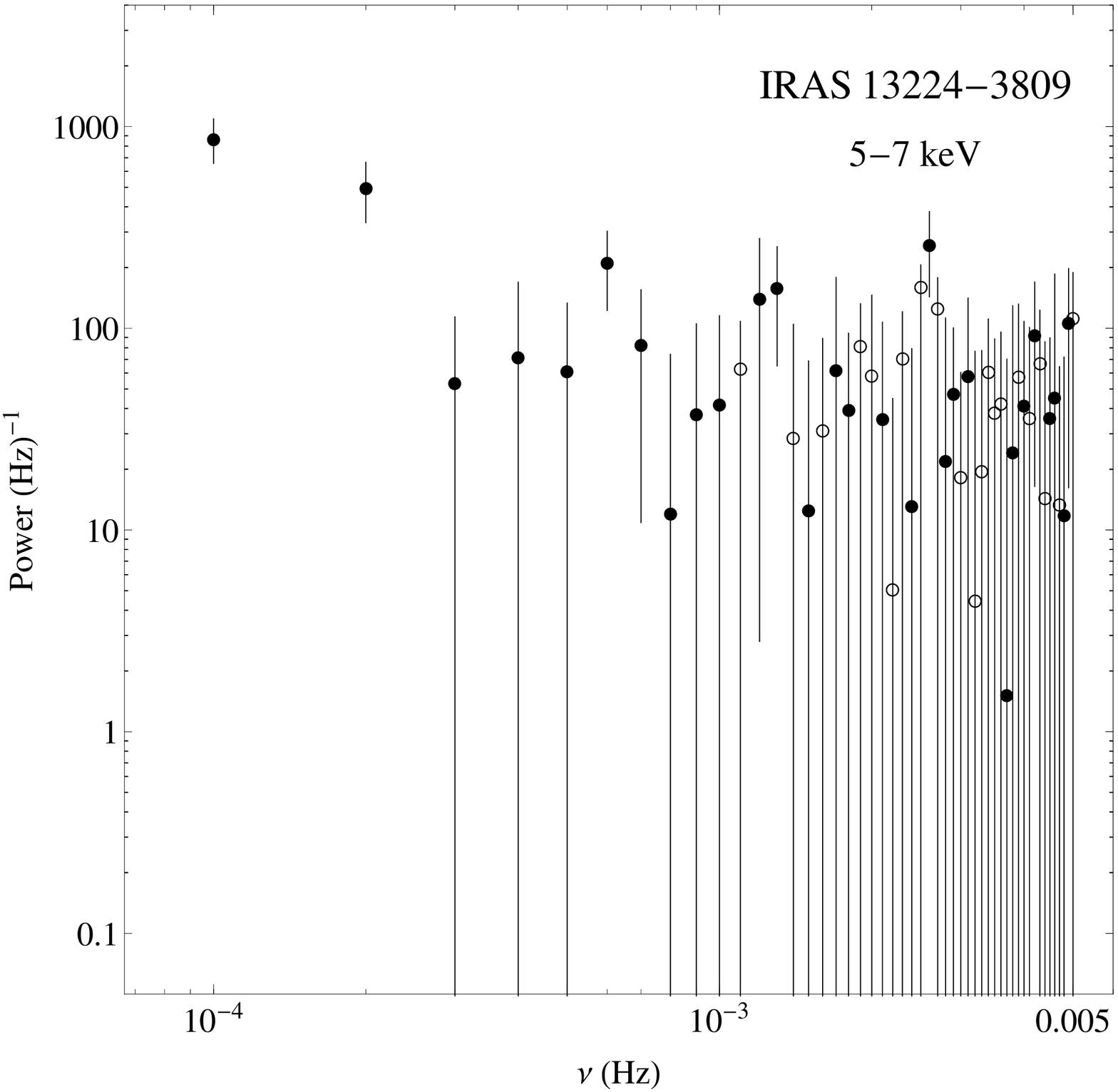}
\caption{The observed PSDs for MCG--6-30-15 and IRAS\;13224-3809 with the Poison noise level (equation~\ref{eqe:poisNoise}) subtracted, in the top and bottom panels, respectively. The filled circles correspond to the positive estimates and the open circles correspond to the absolute values of the negative estimates, i.e.\ when the Poisson noise level, at a given Fourier frequency, is greater than the actual observed PSD value. Left-hand panels: The observed PSDs in the soft-band. Right-hand panels: The observed PSDs in the iron-line band.}
\label{fig:softHardExample}
\end{figure*}

The standard method for the PSD estimation is by calculating the periodogram of a light curve \citep[e.g.][]{priestley81}. For all the sources in our sample, we divide each individual light curve into uninterrupted segments of 10 ks duration. For each segment we compute the periodogram as the modulus-squared of the discrete Fourier transform 
\eqb
P(\nu_j)=\frac{2\Delta t}{N\bar{x}^2}\sum_{i=1}^{N}|x(t_i){\rm e}^{2{\rm\pi i} \nu_j)}|^2
\label{eqe:prdgr}
\eqe
\noindent
where $N\;(=100)$ is the number of points in the 10 ks segment, $\Delta t\;(=100$ s) is the bin size, and $\bar{x}$ is the mean count rate of each segment. The periodogram is estimated at the Fourier frequencies $\nu_j=j/(N\Delta t)$ with $j=1,2,\ldots N/2$, and it is normalized in such a way so that its integral, over the entire frequency range from $1/(N\Delta t)$ to $1/(2\Delta t)$, yields the fractional variance of each segment. We also estimate the Poisson noise level according to the relation

\eqb
PN = \frac{2\overline{\sigma^2_{\rm err}}\Delta t }{\bar{x}^2}
\label{eqe:poisNoise}
\eqe
where  $\overline{\sigma^2_{\rm err}}$ is the mean square error of the points in each segment \citep[e.g.][]{papadakis95}. We subtract this noise level from the individual periodograms, we average them, and we accept the average periodograms as the estimate of the `observed' PSD for each source, in the $0.5-1$ and $5-7$ keV energy bands. 

This PSD estimation method is meaningful in the case when the intrinsic PSD does not vary with time, i.e.\ the X-ray variability process is stationary. The well known rms-flux relation in AGN \citep{uttley01} suggests that AGN PSDs should vary in amplitude with flux and, at least in the case of NGC\;4051, this has been observed \citep{vaughan11}. Even if that is the case in all AGN, the PSDs we estimate should be representative of the mean PSDs in AGN. We discuss this issue further in the Appendices~\ref{app1:nonStation} and \ref{app2:nonStationPSDresid}.

Figure~\ref{fig:softHardExample} shows the observed power spectra of MCG--6-30-15 and IRAS\;13224-3809 (upper and lower panels, respectively). The left and right panels show the soft and iron-line band PSDs, respectively, and the open circles, at high-frequencies, indicate the absolute value of the negative PSD points, i.e.\ when the Poisson noise level, at a given Fourier frequency, is greater than the actual observed PSD value. The iron-line band PSDs at high-frequencies are more `noisy' than the soft-band PSDs. In fact, the iron-line band PSD in the case of IRAS\;13224-3809 is heavily dominated by the experimental noise variations even up to the lowest frequencies probed by the 10 ks segments. As a result, the error of the averaged periodograms at almost all frequencies is very large. This is mainly due to the fact that this source is not as bright as MCG--6-30-15 in the $5-7$ keV band. 

The $5-7$ keV PSD of IRAS\;13224-3809 is characteristic of most sources in the sample. The Poisson noise dominates the observed variations at frequencies higher than $10^{-3}$ Hz, except from four sources, namely MCG--6-30-15, NGC\;4051, Ark\;564, and NGC\;7314. The iron-line band PSD in these sources is well defined over the lowest sampled frequency decade, i.e.\ between $10^{-4}$ and $10^{-3}$ Hz. This is not surprising, as these sources are among the X-ray brightest in our sample (hence the Poisson noise level is low) and they probably host a low mass BH as well, implying large amplitude variations at high-frequencies.

%%%%%%%% FIGURE 2
\begin{figure*}
\includegraphics[width=3.3in]{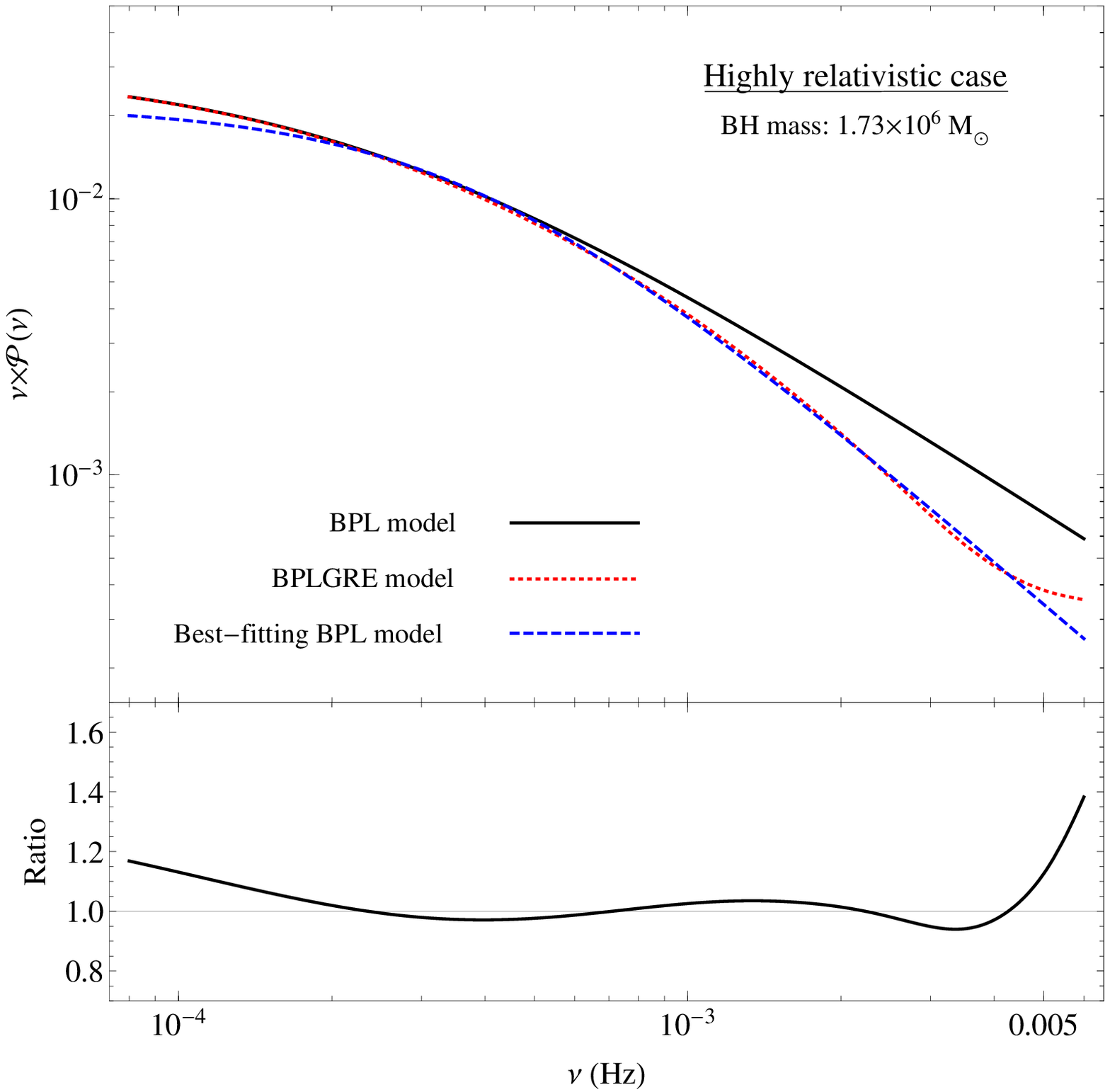}
\includegraphics[width=3.3in]{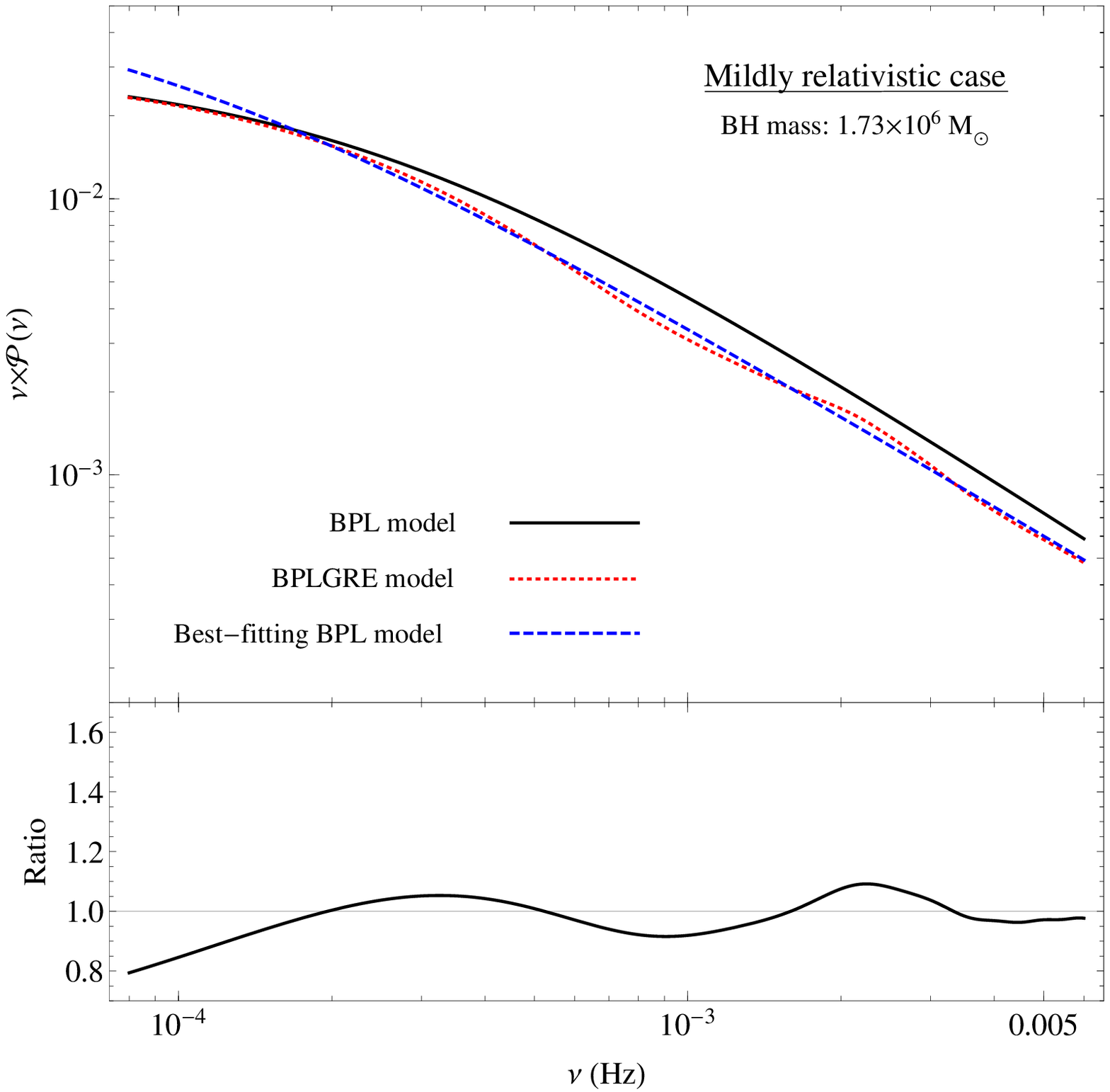}
\includegraphics[width=3.3in]{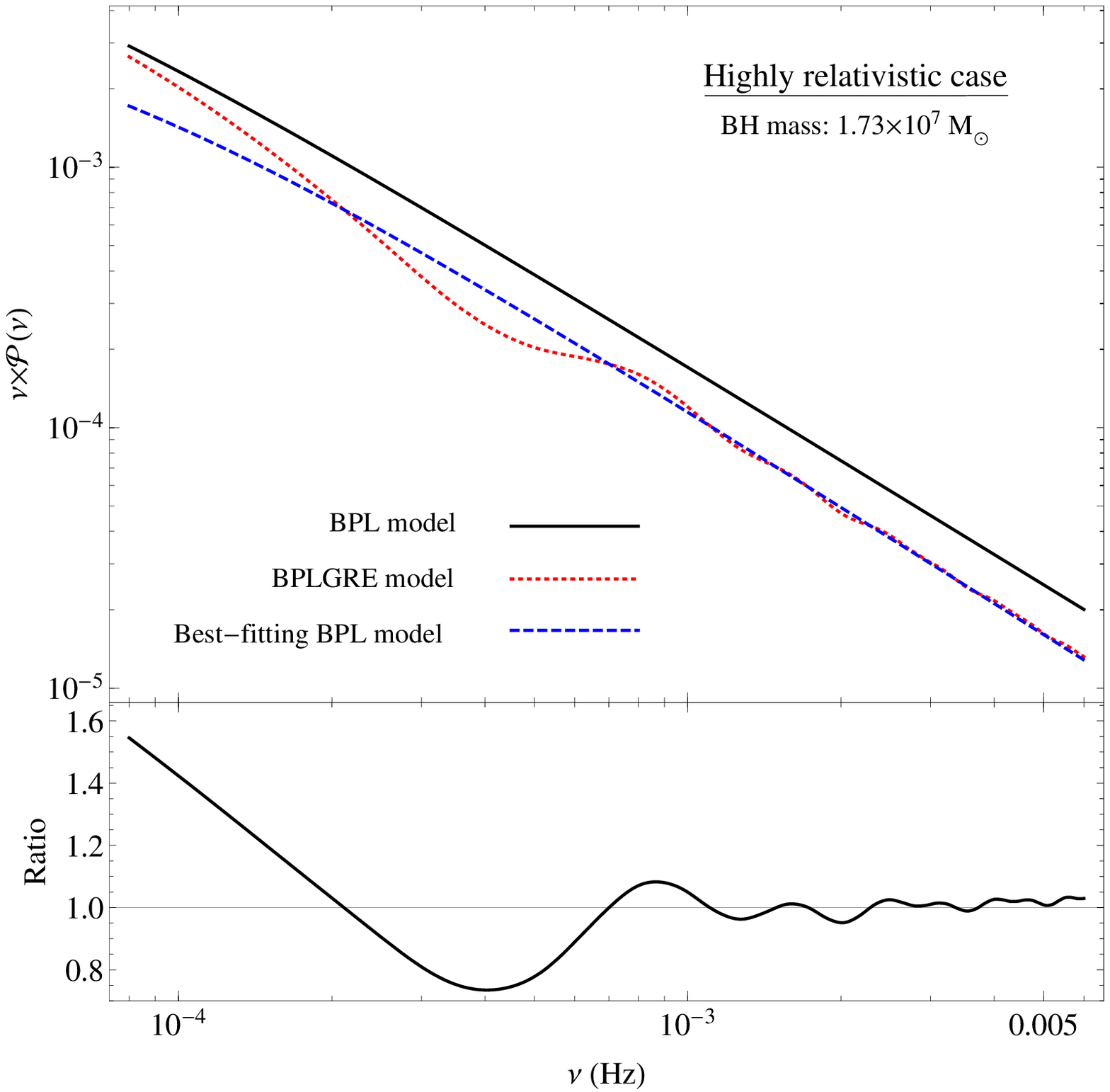}
\includegraphics[width=3.3in]{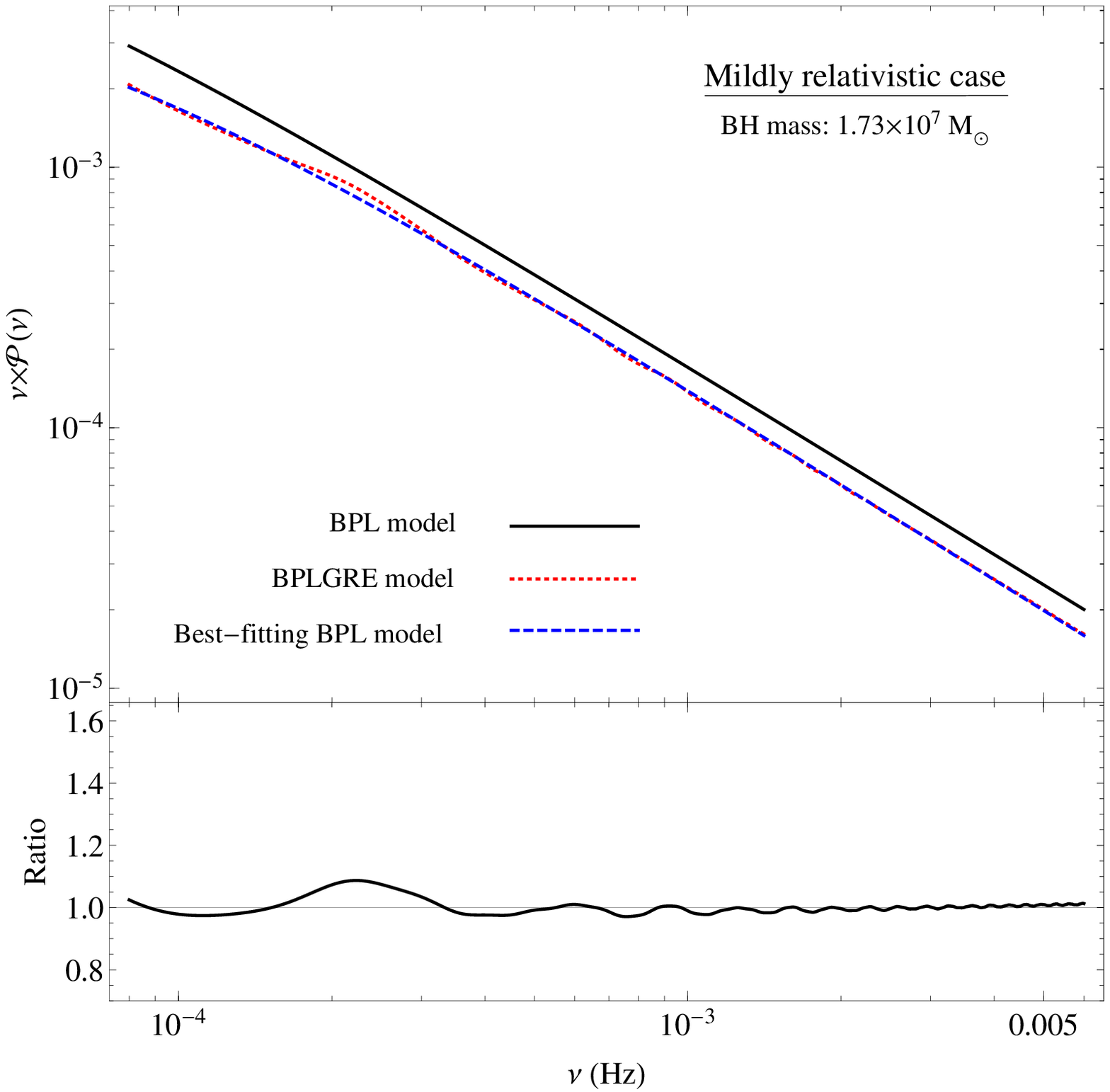}
\caption{The expected GRE features for a BH mass of $1.73\times 10^6$ \ms\ (as the one in NGC\;4051) and $1.73\times 10^7$ \ms, in the top and bottom panels, respectively. Left-hand panels: The `highly relativistic' case ($\alpha=1$, and a point-like X-ray source located at a height of $h=3.6$ \rg). The black solid lines indicate the BPL model PSD, the red dotted lines indicate the respective BPLGRE model PSDs and the blue dashed lines indicate the best-fitting BPL model to the BPLGRE model PSDs. The attached plots, at the bottom of each panel, show the ratio of the BPLGRE model PSDs (red dotted lines in the upper panels) over the respective best-fitting BPL model (blue dashed lines in the upper panels). Right-hand panels: The same as before for a `mildly relativistic' case ($\alpha=0$ and $h=21.3$ \rg) (see Sect.~\ref{ssect:objective2} for further details).}
\label{fig:models}
\end{figure*}

%%%%%%%%%%%%%%%%%%%%%%%%%%%%%%%%%%%%%%%%
\subsection{Our objective}
\label{ssect:objective1}
%%%%%%%%%%%%%%%%%%%%%%%%%%%%%%%%%%%%%%%%
Initially, our plan is to fit the observed PSDs with the so-called bending power law (BPL) model \citep[e.g.][]{mchardy04}
\noindent
\eqb
\mathscr{P}(\nu_j;\mathbf{\mathit \gamma})=A\nu_j^{-1}\left[1+\left(\frac{\nu_j}{\nu_{\rm b}}\right)^{s_{\rm h}-1}\right]^{-1}
\label{eq:bpl}
\eqe
\noindent
describing a PSD model with a slope of $-1$ at low-frequencies, bending gradually to a slope of $s_{\rm h}$ at high-frequencies above the bend-frequency, $\nu_{\rm b}$. This model provides a good representation of the X-ray PSDs of many AGN \citep[e.g.][]{gonzalez12}. The vector $\mathbf{\mathit\gamma}=\{A,\nu_{\rm b}, s_{\rm h}\}$ defines the fitting parameter space, having as components the PSD model parameters. Note that since we freeze the low-frequency slope to -1, it is always possible to derive a best-fitting value for the bend-frequency, $\nu_{\rm b}$, even if it is actually located at a frequency lower than $10^{-4}$ Hz (i.e.\ the lowest frequency in the observed PSDs). Consider for example the case of a high BH mass AGN having the PSD bend-frequency around $10^{-5}$ Hz and for which the best-fit $s_{\rm h}$ is equal to 2 (as derived from the model fit at frequencies greater than $10^{-4}$ Hz.). In this case, the minimization routine will yield a value for the best-fitting bend-frequency well below 
$10^{-4}$ Hz, but the best-fitting $\nu_{\rm b}$ estimate will have huge uncertainties. The upper error on this model parameter will still be meaningful, but the lower limit on the error will be so large that in effect the bend-frequency estimate will set an upper limit on $\nu_b$, thus dictating the poor quality of this parameter estimate.

According to P16, if X-ray reprocessing takes place in the innermost parts of the accretion disc, the observed PSDs should be equal to the product of the intrinsic power spectrum (i.e.\ the PSD due to the X-ray variability mechanism itself)  and the square of the transfer function, i.e.\ the Fourier transform of the disc's response function. In this case, if the intrinsic PSD has a BPL shape, the observed PSD should be given by a model of the form  

\eqb
\mathscr{P}(\nu_j;\mathbf{\mathit\gamma})=A\nu_j^{-1}\left[1+\left(\frac{\nu_j}{\nu_{\rm b}}\right)^{s_{\rm h}-1}\right]^{-1}\times\left|\Gamma(\nu_j;h,\alpha,\theta)\right|^2
\label{eq:bplgre}
\eqe
where $\Gamma(\nu;h,\alpha,\theta)$ is the transfer function (see equation~5 in P16). This is a bending power law model which is modified by the GRE component (BPLGRE model, hereafter) which holds all the important information regarding the relativistic echo features in the case of X-ray reflection. For a given BH mass, the function $\Gamma(\nu;h,\alpha,\theta)$ depends on the height of the X-ray source, $h$, the BH spin, $\alpha$, and the inclination angle, $\theta$. Vector $\mathbf{\mathit\gamma}=\{A,\nu_{\rm b}, s_{\rm h},h,\alpha,\theta\}$ defines the fitting parameter space in this case. We note that the normalization, $A$, in the equation above is not identical to the normalization $A$ of the BPL model. In addition of being representative of the BPL component normalization, $A$ also accounts for the normalization of the `General Relativistic Echo' (GRE) component, i.e.\ $\left|\Gamma(\nu_j;h,\alpha,\theta)\right|^2$, as well. The normalization of the latter component will be equal to the one fitted in 
the $5-7$ keV energy band PSDs (as the P16 transfer functions were estimated for that band), if the conditions are identical to the ones described in P16 (i.e.\ neutral disc and Solar iron abundance). In all other cases, we expect the overall shape of the transfer functions to be similar to those estimated by P16, but having smaller or larger amplitude variations, hence the presence of a different than unity normalization for the GRE component, which is included within $A$, given the way we have defined the BPLGRE model in equation~\ref{eq:bplgre}. 

After fitting the simple BPL model (equation~\ref{eq:bpl}) to the observed PSDs, we will also fit to them the BPLGRE model (equation~\ref{eq:bplgre}). The main objective of this study is to compare the goodness of fit between the BPL and BPLGRE models and investigate whether the BPLGRE is favoured over the BPL model. In such a scenario reprocessing GRE signatures should be imprinted in the X-ray PSDs of AGN, despite the fact that such detections are not trivial, as we discuss below.

\subsection{Detecting the GRE signatures in the currently observed PSDs}
\label{ssect:objective2}

In this section we check how `easy it will be to detect the GRE features in practice. The continuous lines in the upper two panels of Fig.~\ref{fig:models} show the BPL model with $A=0.03$, $\nu_{\rm b}=2.3\times10^{-4}$ Hz, and $s_{\rm h}=2.2$ \citep[these parameters are similar to the best-fitting BPL model parameters in the $0.5-1$ keV band periodogram of NGC\;4051;][]{emmanoulopoulos13b}. The continuous lines in the lower panels of the same figure show the BPL model in the case of the PSD bend-frequency being ten times lower. The BPL normalization is defined in such a way so that the power at the bend-frequency (i.e.\ PSD$(\nu_b)\times \nu_b$) will be the same in both PSDs. In all panels, the data are plotted in the frequency range that our PSD samples are estimated (i.e.\ between $10^{-4}$ and $5\times 10^{-3}$ Hz).

The red dotted lines indicate the same BPL models multiplied by $|\Gamma(\nu;3.6\;r_{\rm g},1,40\degr)|^{2}$ and $|\Gamma(\nu;21.3\;r_{\rm g},0,40\degr)|^{2}$ (left and right panels, respectively). The former is the square of the transfer function in a `highly relativistic' case: a Kerr BH where the lamp is located at a distance of $3.6$ \rg\; above the disc \citep[this is similar to the mean height of the X-ray source in AGN, according to the results of][]{emmanoulopoulos14}. The latter corresponds to a `mildly relativistic' case: a non-rotating BH and the X-ray source located further away from the disc at 21.3 \rg. In the upper panel plots, we assume $M_{\rm BH}=1.73\times 10^6$ \ms, i.e.\ the BH mass estimate of NGC\;4051. In the lower panels we assume a ten times larger BH mass (since the bend-frequency is ten times lower). If the intrinsic PSDs have a BPL shape, and the X-ray/disc geometry in AGN is similar to the lamp-post configuration, the observed PSDs should be similar to the red dotted lines, i.e.\
 the expected BPLGRE PSDs, in Fig.~\ref{fig:models}. 

To check in practice the detectability of the GRE features, we fit a simple BPL model (black solid lines) to the predicted BPLGRE PSDs (red dotted lines). The resulting best-fitting models are indicated by the blue dashed lines. The bottom plots in each panel indicate the ratio of the BPLGRE PSDs over the best-fitting BPL models to them. Interestingly, a simple BPL model fits well the BPLGRE model PSDs over the frequency range we sample. The GRE features have a smaller amplitude in the `mildly relativistic' case but, even in the `highly relativistic' case, the best-fitting ratios are less than 20 per cent, at all frequencies, except from the lowest frequency point in the large BH mass case (see the bottom plot in the lower left-hand panel of Fig.~\ref{fig:models}).

The results above show that it will be difficult to detect the GRE effects in objects with small BHs (like NGC\;4051). The observed PSDs should be be well fitted by a BPL model, with a slope steeper than the intrinsic one (which is unknown a priori). The amplitude of the expected best-fitting residuals for a BPL model fit will be less than 20 per cent, even in the `highly relativistic' case. The amplitude of the residuals may increase at the highest frequency sampled by the data, but the PSD uncertainties due to the Poisson noise will be much larger there, even for the brightest and most variable sources in our sample. Similarly, the detection will not be much easier for larger BH mass objects. The largest amplitude residual is expected at the lowest frequency points in the observed PSDs in the `highly relativistic' case. At all the other frequencies the residuals' amplitude will be smaller than 20 per cent, and significantly smaller in the `mildly relativistic' case. 

The main objective of the discussion above is not to provide an extensive investigation of the detectability of the GRE features in the observed PSDs, taking into account the full range of the BH masses in the objects of the sample and the count rate of the available light curves. It mainly aims to highlight the difficulties associated with the detection of the GRE effects in the {\it individual} PSDs: for most objects in the sample, and for most of the observed frequency range, the error of the observed PSDs is larger than 20 per cent. With this in mind, we present below the best-fitting results of the BPL and BPLGRE models to the observed PSDs.

%%%%%%%%%%%%%%%%%%%%%%%%%%%%%%%%%%%%%%%
\section{PSD MODEL FITTING}
%%%%%%%%%%%%%%%%%%%%%%%%%%%%%%%%%%%%%%%
\label{sect:PSD_modelFit}
The model fitting is performed using the traditional $\chi^2$ as the fit statistic. For the minimization of the $\chi^2$ merit function, we use the classical Levenberg-Marquardt method \citep{bevington92}. The number of light curve segments is larger than 50 for 8 out 12 sources in our sample. In this case, the distribution of the averaged periodogram estimates approximates reasonably well a Gaussian distribution \citep{papadakis93}. We therefore assume Gaussian statistics, and we quote the 68.3 per cent confidence intervals for the each best-fitting parameter, estimated by increasing $\Delta\chi^2$ by one. We fit models to the iron-line PSDs of the four sources that we discussed above (i.e.\ the ones with well defined, low-frequency shapes), and to the soft-band PSDs of all the sources in our sample. 

%%%%%%%%%%%%%%%%%%%%%%%%%%%%%%%%%%%%%%%
\subsection{Bending power law (BPL) model fits}
%%%%%%%%%%%%%%%%%%%%%%%%%%%%%%%%%%%%%%%
\label{ssect:bplmodel}

Table~\ref{tab:bplHard} and Table~\ref{tab:bplSoft} list our best-fitting BPL model results for the $5-7$ keV and $0.5-1$ keV observed PSDs, respectively. Our best fit parameter values are similar to those usually determined by the BPL model fits to AGN power spectra. For example, our soft-band, best-fitting $\nu_b$ and $\alpha$ values are entirely consistent (within the errors) with the $0.2-2$ keV, Model B fit results of \citet{gonzalez12} for the same sources.  
 
\begin{figure*}
\includegraphics[width=3.3in]{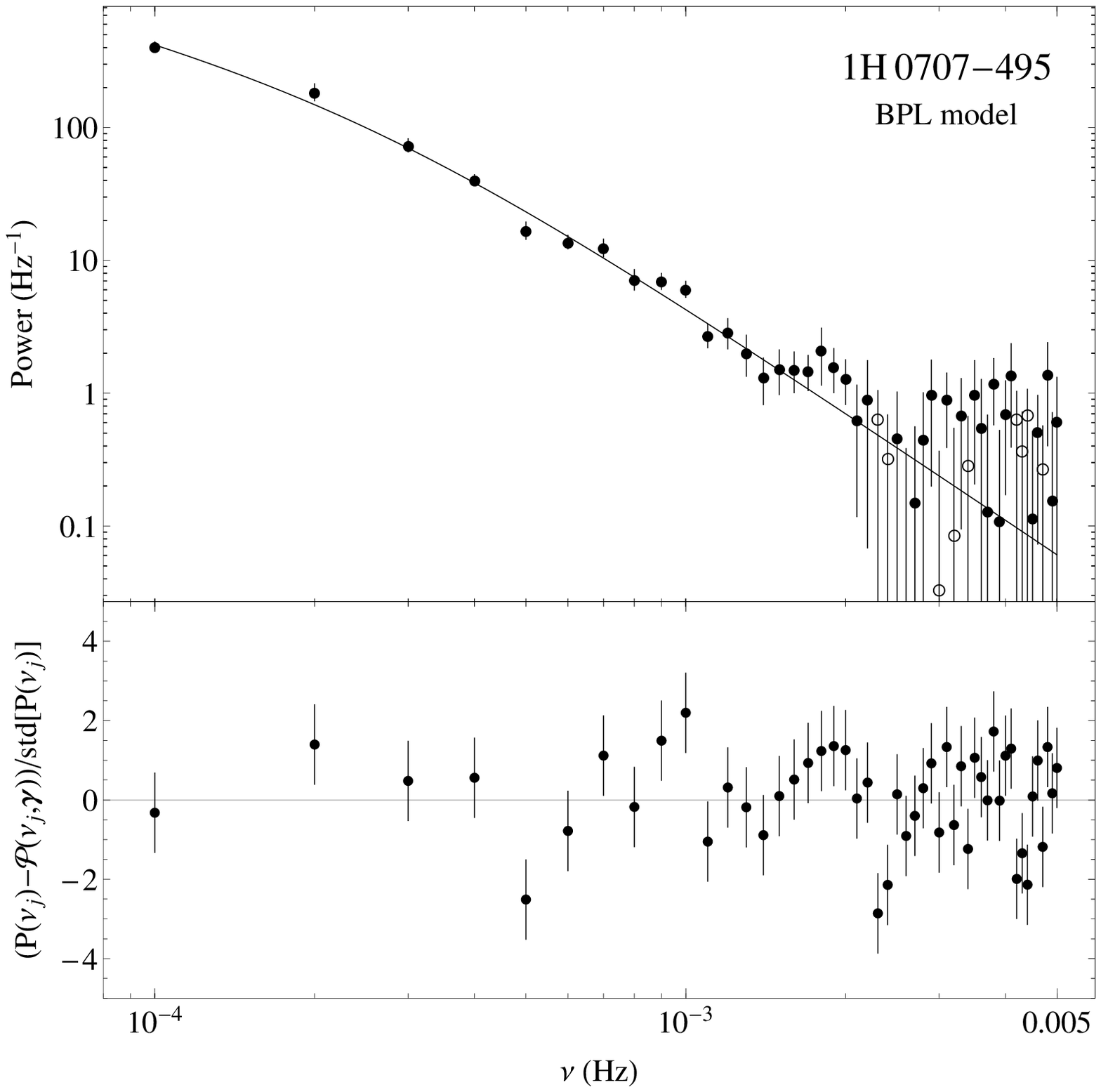}\hspace{1em}
\includegraphics[width=3.3in]{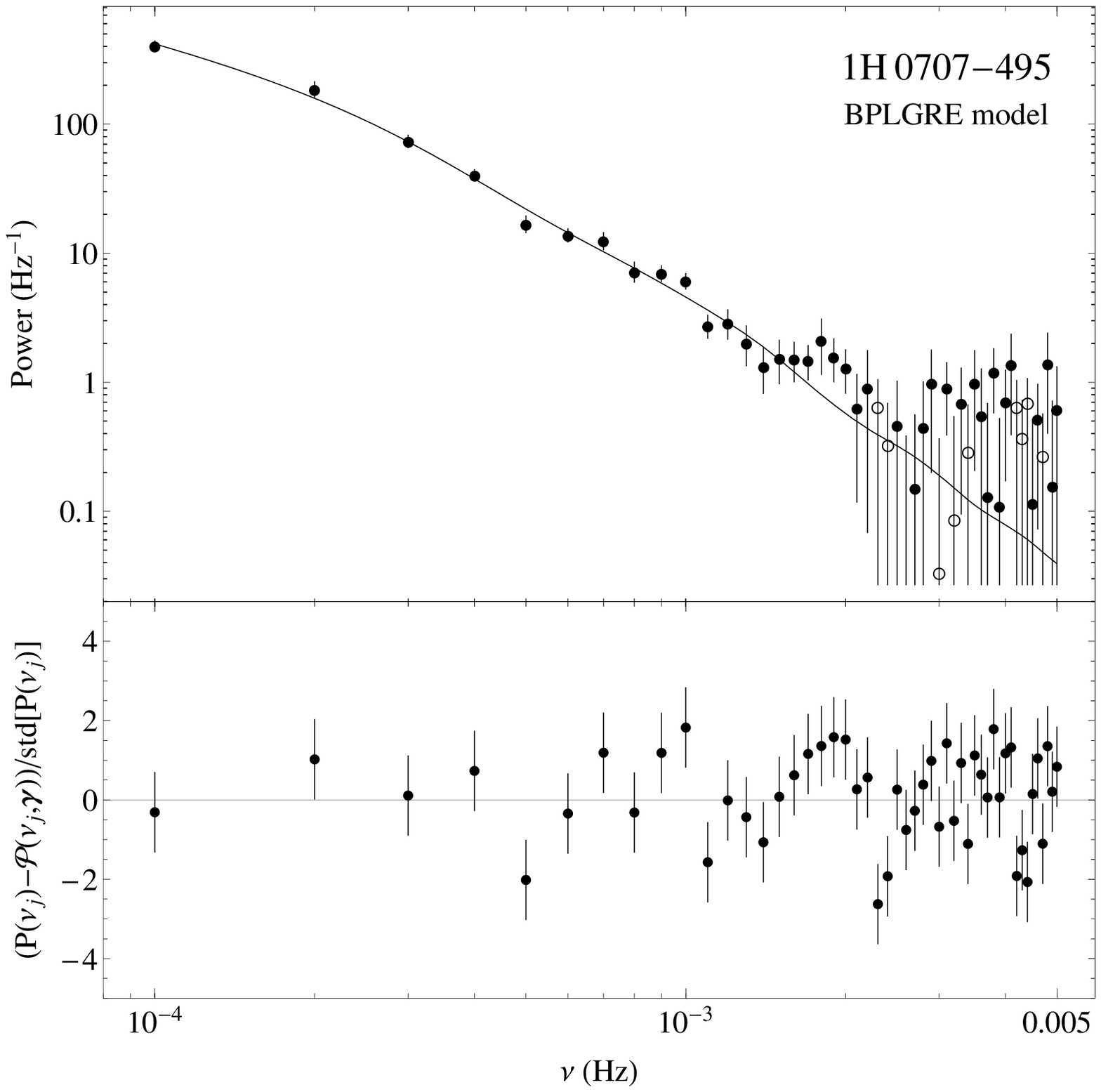}
\includegraphics[width=3.3in]{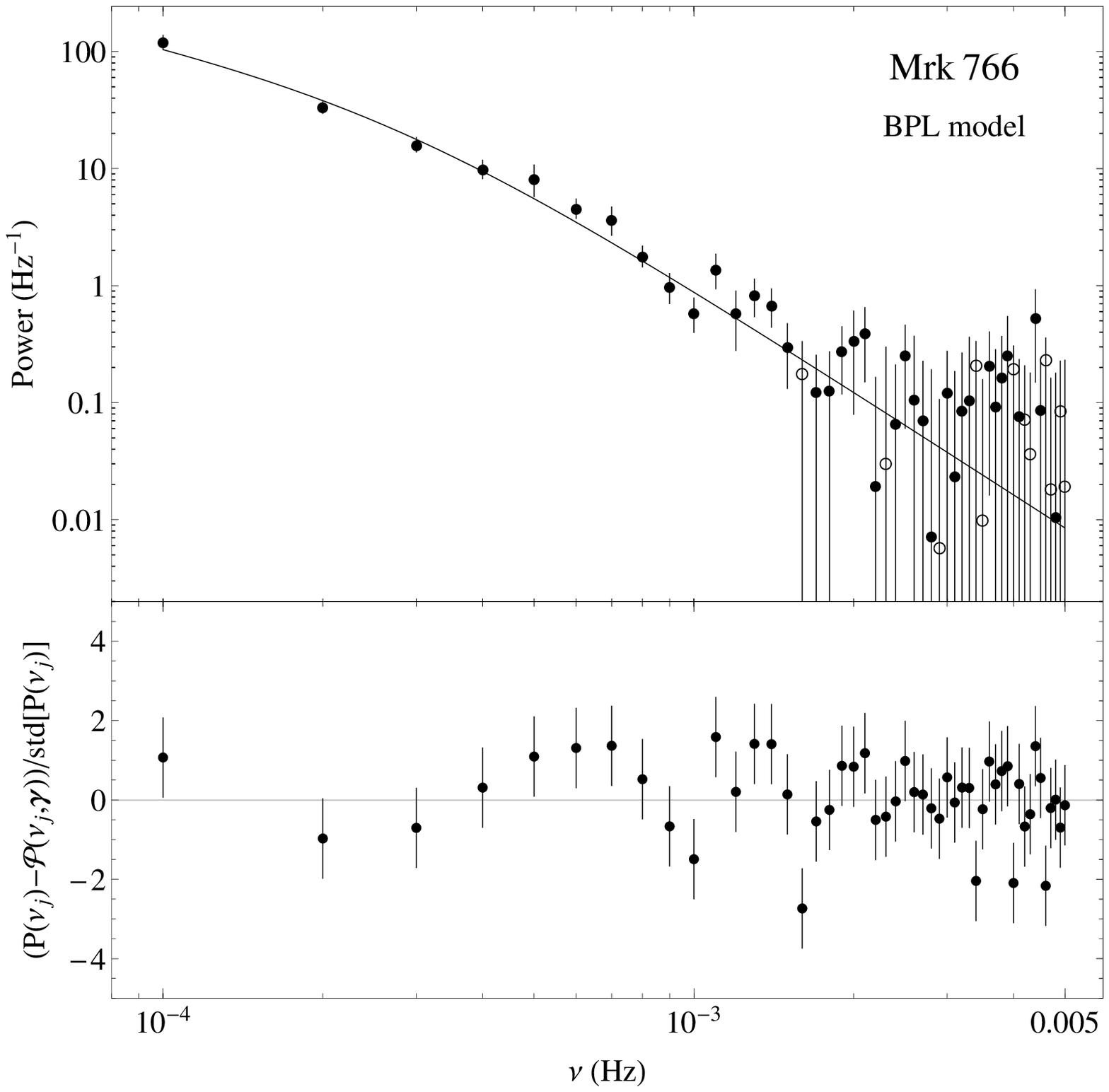}\hspace{1em}
\includegraphics[width=3.3in]{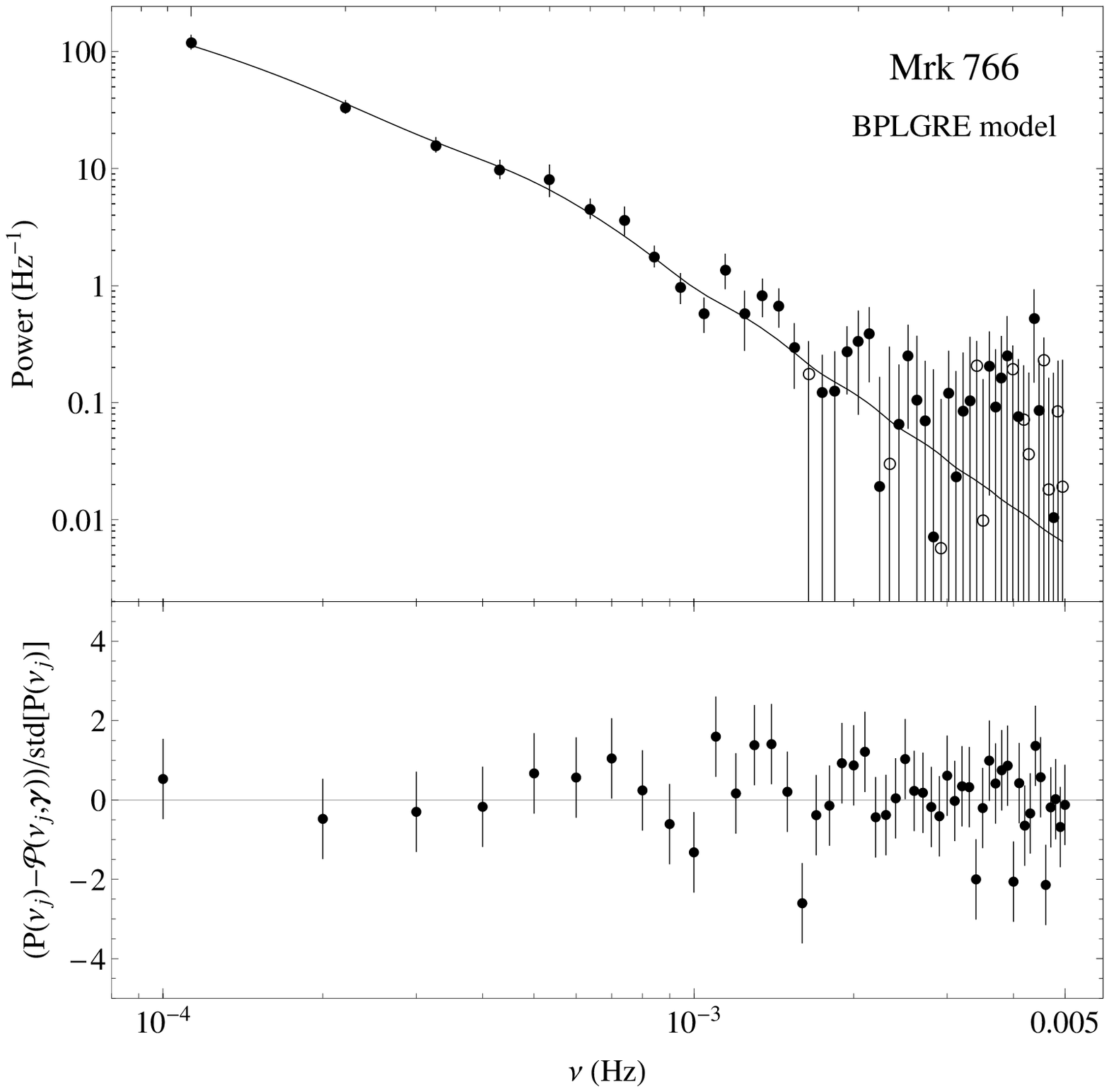}
\caption{The best-fitting PSD models in the soft-band (solid lines) for 1H\;0707-495 and Mrk\;766 (top and bottom panels, respectively) together with the corresponding best-fitting residual plots attached at the bottom of each panel. The filled circles correspond to the positive PSD estimates and the open circles correspond to the absolute values of the negative PSD estimates, i.e.\ when the Poisson noise level, at a given Fourier frequency, is greater than the actual observed PSD value. Left-hand panels: The best fitting BPL model. Right-hand panels: The best-fitting BPLGRE model.}
\label{fig:BestFitExamples}
\end{figure*}

The left-hand panels in Fig.~\ref{fig:BestFitExamples} show the best-fit BPL models to the soft-band PSDs of 1H\;0707-495 and Mrk 766 (top and bottom panels, respectively). The lower plots, attached to each panel, show the best-fitting residual at each Fourier frequency, $\nu_j$, i.e.\ $[\mathscr{P}_{\rm best-fit}(\nu_j;\mathbf{\mathit\gamma})-P_{\rm obs}(\nu_j)]/\std[P_{\rm obs}(\nu_j)]$ in which $\std[P_{\rm obs}(\nu_j)]$ is the standard deviation of the average periodograms. According to the resulting $\chi^2$ values, the BPL model fits well the observed power spectrum of Mrk\;766, but less so the 1H\;0707-495 PSD (the null hypothesis probability, $p_{\rm null}$, is less than 2.2 per cent in this source). In fact, the null hypothesis probability for the best-fitting BPL model fit to the iron-line PSDs of NGC\;7314 and Ark\;564 is rather low ($p_{\rm null}$=1.6 and 0.6 per cent, respectively). $p_{\rm null}$ is also low in the case of the BPL best-fits to the soft-band PSDs of SWIFT\;J2127+5654 and NGC\;
3516 ($p_{\rm null}=2.1$ and $0.3$ per cent, respectively). If the BPL model would represent correctly the underlying PSD of all sources, we would expect just one or two $p_{\rm null}$ values smaller then 5 per cent among the 16 fitted PSDs, as opposed to the 5 cases among our sample. We discuss in Appendix~\ref{app2:nonStationPSDresid} the case of the `bad' BPL model fits in more detail.

\begin{table}
 \caption{Bending power law model fit results, $5-7$ keV band PSDs (for 47 d.o.f.).}
 \label{tab:bplHard}
 \begin{tabular}{lcccc}\hline
  & \multicolumn{3}{c}{Bending power law ($5-7$ keV)} & \\ \hline
AGN name		& $A$		& $\nu_{\rm b}$			& 	$s_{\rm h}$	& $\chi^2$	\\ 
			& ($10^{-3}$ Hz$^{-1}$)		&  ($10^{-4}$ Hz)	&			&			\\
\hline
NGC\;7314		& $12.7^{+4.1}_{-3.8}$		& $3.82^{+1.59}_{-1.88}$	& $2.84^{+0.43}_{-0.61}$  	& $70.1$    \\
NGC\;4051		& $39.2^{+15.2}_{-17.3}$    	& $1.87^{+1.13}_{-0.94}$	& $2.29^{+0.20}_{-0.29}$	& $49.5$		\\
Ark\;564		& $12.8^{+2.3}_{-2.4}$	    	& $6.64^{+1.95}_{-1.83}$	& $3.59^{+1.12}_{-1.32}$   	& $75.6$      \\
MCG--6-30-15		& $9.4^{+3.2}_{-2.5}$	    	& $2.91^{+1.46}_{-1.08}$	& $2.59^{+0.29}_{-0.32}$    	& $53.1$      \\
\hline
\end{tabular}
\end{table}

\begin{table}
 \caption{Bending power law model fit results, $0.5-1$ keV band PSDs (for 47 d.o.f.).}
 \label{tab:bplSoft}
 \begin{tabular}{lcccc}\hline
  & \multicolumn{3}{c}{Bending power law ($0.5-1$ keV)} & \\ \hline
AGN name		& $A$		& $\nu_{\rm b}$			& $s_{\rm h}$		& $\chi^2$	\\ 
			&  ($10^{-3}$ Hz$^{-1}$)		&  ($10^{-4}$ Hz)	&				&			\\
\hline
NGC\;7314		    & $15.9^{+6.5}_{-7.2}$		    & $2.89^{+0.18}_{-0.19}$	& $2.47^{+0.52}_{-0.54}$  	& $51.8$    \\
NGC\;4051		    & $111^{+37}_{-30}$    & $1.39^{+0.59}_{-0.62}$	& $2.24^{+0.08}_{-0.11}$	& $58.4$		\\
Mrk\;766		    & $11.9^{+2.3}_{-2.2}$	    & $2.68^{+0.66}_{-0.57}$	& $2.93^{+0.23}_{-0.20}$    & $48.9$		\\
Ark\;564		    & $5.9^{+0.7}_{-0.7}$	    & $6.79^{+0.10}_{-0.13}$	& $2.89^{+0.17}_{-0.14}$    & $54.8$		\\
1H\;0707-495	    & $51.4^{+8.3}_{-7.9}$	    & $2.41^{+0.48}_{-0.42}$	& $2.69^{+0.15}_{-0.16}$	& $68.5$		\\
MCG--6-30-15		& $23.4^{+8.7}_{-6.3}$	    & $1.36^{+0.52}_{-0.42}$	& $2.49^{+0.21}_{-0.12}$    & $54.2$		\\
IRAS\;13224         & $531^{+424}_{-295}$	& $0.16^{+0.60}_{-\text{---}}$    &$2.10^{+0.13}_{-0.11}$	    & $42.6$    \\
\hspace{2.8em}-3809    &               \\
SWIFT\;J2127        & $10.3^{+5.4}_{-6.3}$	    & $1.83^{+1.25}_{-1.32}$	& $3.2\pm1.3$             & $68.4$ 		\\
\hspace{3.34em}+5654    &               \\               
Mrk\;335 		    & $4.3\pm1.9$ 		        & $2.97^{+1.23}_{-1.09}$	& $3.42^{+0.62}_{-0.76}$  	& $48.2$	\\
NGC\;3516		    & $7.1^{+6.1}_{-7.2}$	    & $0.48^{+0.49}_{-0.46}$	& $2.51^{+0.62}_{-0.56}$    & $78.2$      \\
PG\;1211+143        & $8.6^{+2.3}_{-4.9}$        & $0.61^{+0.45}_{-0.56}$    & $2.94^{+0.34}_{-0.73}$    & $61.7$    \\
PKS\;0558-504		& $11.7^{+8.9}_{-7.9}$	    & $0.58^{+0.44}_{-0.39}$	& $2.38^{+0.31}_{-0.36}$ 	& $21.8$        \\
\hline
\end{tabular}
\end{table}

%%%%%%%%%%%%%%%%%% FIGURE 3

\subsection{Bending power law with general relativistic echo (BPLGRE) model fits}

In order to define the BPLGRE models that we fit to the observed PSDs, we compute the transfer function as explained by P16, assuming three values for the BH spin: $\alpha=0$ (Schwarzschild BH, $r_{\rm in}= 6$ \rg), 0.676 (intermediate spin BH, $r_{\rm in}= 3$ \rg), and 1 (Kerr BH, $r_{\rm in}= 1$ \rg). Regarding the height of the X-ray source, we consider a variety of values, depending on the BH spin, 18 values in the case of the Schwarzschild BHs (2.3, 2.9, 3.6, 4.5, 5.7, 7, 8.8, 11, 13.7, 17.1, 21.3, 26.5, 33.1, 41.3, 51.5, 64.3, 80.2, and 100 \rg), we add the value of $1.9$ \rg\ in the case of BHs with $\alpha=0.676$, and the value of 1.5 \rg\ in the case of Kerr BHs. As for the inclination angle, we consider systems that are observed by a distant observer at a viewing angle of $\theta=20, 40$ and $60$ degrees ($\theta=0$ or 90 degrees correspond to a disc seen face on and edge on, respectively). In total, the model parameter space, regarding the physical quantities $\alpha, h,$ and $\theta$, yields a 
total of $(20+19+18)\times 3 = 171$ different geometrical layouts of the lamp-post model.

In principle, M$_{\rm BH}$ should also be left as a free parameter during the model fitting process. However, given the complex nature of the fitting procedure, and the need to determine as accurately as possible the model parameters, we compute the 171 model transfer functions assuming the M$_{\rm BH}$ estimates listed in Table~\ref{tab:obs}. Note that we actually repeated the fits by letting M$_{\rm BH}$ be a free model parameter; we comment on these results at the end of this section. Then, using the 171 model transfer functions, we create a linearly interpolated function of $\left|\Gamma(\nu;h,\alpha,\theta)\right|^2$, describing the entire parameter space among the 171 transfer function models. This interpolated transfer function is used in equation~\ref{eq:bplgre} to fit the observed PSDs. 

\begin{table*}
 \caption{Bending power law with General Relativistic echo model best-fit results,  $5-7$  keV  band
PSDs (for 44 d.o.f.).}
 \label{tab:bplGRhard}
 \begin{tabular}{lccccccc}
  \hline
  & \multicolumn{3}{c}{Bending power law model} & \multicolumn{3}{c}{GR Echo model} & \\ \hline
AGN name	& $A$	              & $\nu_{\rm b}$		& $s_{\rm h}$	&  $h$ 	& $\theta$ 	&	$\alpha$	& $\chi^2$	 \\
			&  ($10^{-3}$ Hz$^{-1}$)	&  ($10^{-4}$ Hz)		& 		&  (\rg)	& $(\degr)$	&	&			\\
\hline
NGC\;7314		& $12.6^{+3.0}_{-2.6}$       & $5.36^{+1.91}_{-1.42}$	& $3.37^{+0.83}_{-0.69}$	& $64.91^{+10.12}_{-11.43}$     &  
$37.81^{+11.31}_{-12.78}$       & $0.82^{+0.15}_{-0.29}$        & $68.9$    		\\
NGC\;4051       & $32.3^{+8.4}_{-6.3}$		& $3.62^{+1.10}_{-1.18}$	& $2.60^{+0.32}_{-0.21}$	& $52.34^{+15.27}_{-12.46}$	    &       $27.74^{+11.46}_{-\text{---}}$  & $0.89^{+\text{---}}_{-0.42}$  & $45.8$ \\
Ark\;564        & $15.9^{+4.6}_{-2.5}$		& $6.93^{+2.03}_{-1.54}$	& $3.23^{+1.25}_{-1.24}	$	& $5.5^{+6.52}_{-\text{---}}$	&       $35.35^{+16.87}_{-13.39}$	    & $0.51^{+0.42}_{-0.38}$		& $73.7$  \\ 
MCG--6-30-15    & $10.0^{+10.4}_{-9.4}$       & $4.06^{+0.86}_{-0.75}$	& $2.99^{+0.37}_{-0.42}$	& $27.43^{+21.31}_{-19.23}$	    &       $46.06^{+13.12}_{-12.54}$       & $0.92^{+\text{---}}_{-0.28}$	& $47.7$ \\
\hline
\end{tabular}
\end{table*}

\begin{table*}
 \caption{Bending  power law with General Relativistic echo model best-fit results, $0.5-1$  keV  band
PSDs (for 44 d.o.f.).}
 \label{tab:bplGRsoft}
 \begin{tabular}{lcccccccc}
  \hline
  & \multicolumn{3}{c}{Bending power law model} & \multicolumn{3}{c}{GR echo model} & \\ \hline
AGN name	& $A$	              & $\nu_{\rm b}$		& $s_{\rm h}$	&  $h$ 	& $\theta$ 	&	$\alpha$	& $\chi^2$ \\
			&  ($10^{-3}$ Hz$^{-1}$)	&  ($10^{-4}$ Hz)		& 		&  (\rg)	& $(\degr)$	&	&	&		\\
\hline
NGC\;7314			& $14.5^{+4.7}_{-3.8}$      & $4.10^{+2.12}_{-1.98}$	& $2.44^{+0.52}_{-0.49}$	& $43.21^{+12.23}_{-11.02}$	&   $28.43^{13.31}_{-\text{---}}$	& $0.93^{+\text{---}}_{-0.38}$		& $50.1$  
\\
NGC\;4051			& $88.6^{+21.4}_{-19.3}$	& $2.13^{+0.70}_{-0.73}$	& $2.27^{+0.09}_{-0.11}$	& $16.12^{+14.32}_{-10.31}$	& $38.61^{+9.21}_{-8.69}$		& $0.36^{+0.35}_{-0.22}$		& $55.3$    
\\
Mrk\;766			& $12.9^{+2.3}_{-2.2}$		& $2.69^{+0.15}_{-0.14}$	& $2.98^{+0.16}_{-0.17}$	& $30.22^{+20.11}_{-18.51}$	& $24.12^{+9.43}_{-\text{---}}$	& $0.67^{+0.22}_{-0.28}$		& $42.2$    
\\
Ark\;564			& $17.6^{+14.6}_{-11.3}$	& $0.78^{+1.81}_{-\text{---}}$	& $1.71^{+0.18}_{-0.19}	$	& $1.92^{+2.52}_{-\text{---}}$	& $53.49^{+\text{---}}_{-23.39}$	& $0.49^{+0.36}_{-0.28}$		& $49.3$  
\\
1H\;0707-495		& $46.4^{+5.1}_{-7.2}$		& $3.71^{+0.48}_{-0.38}$	& $2.69^{+0.89}_{-0.95}$	& $19.32^{+4.46}_{-6.45}$	& $32.45^{+8.54}_{-9.21}$ 	& $0.16^{+0.42}_{-\text{---}}$	& $64.5$    
\\
MCG--6-30-15		& $33.1^{+9.0}_{-7.8}$		& $1.20^{+0.49}_{-0.38}$	& $2.48^{+0.19}_{-0.26}$	& $25.06^{+14.28}_{-16.96}$	& $40.37^{+11.23}_{-10.31}$	& $0.98^{+\text{---}}_{-0.38}$	& $53.3$ 
\\
IRAS\;13224-3809	& $199^{+164}_{-121}$		& $0.69^{+0.84}_{-0.67}$	& $2.26\pm0.14$			& $16.94^{+7.54}_{-9.44}$	& $25.56^{12.26}_{-\text{---}}$	& $0.54^{+0.34}_{-0.28}$		& $38.6$  
\\
SWIFT\;J2127.4+5645	& $11.9^{+9.1}_{-10.4}$		& $2.00^{+1.30}_{-1.26}$	& $3.40^{+1.04}_{-0.78}$	& $4.23^{+3.21}_{-2.36}$	& $59.81^{+\text{---}}_{-18.23}$	& $0.09^{+3.25}_{-\text{---}}$		& $66.1$   
\\
Mrk\;335			& $9.1^{+3.2}_{-2.6}$		& $2.64^{+0.98}_{-1.06}$	& $3.47^{+0.62}_{-0.58}$	& $3.64^{+5.43}_{-2.21}$	&	$47.66^{+9.32}_{-8.24}$		& $0.91^{+\text{---}}_{-0.27}$	& $46.0$ 
\\
NGC\;3516 			& $2.6^{+2.4}_{-2.2}$ 		& $2.37^{+2.81}_{-1.99}$	& $2.83^{+1.28}_{-1.45}$ 	& $1.52^{+3.25}_{-\text{---}}$	& $23.49^{+12.56}_{-\text{---}}$	& $0.86^{+\text{---}}_{-0.33}$		& $77.1$ 
\\
PG\;1211+143        & $12.5^{+6.5}_{-7.2}$ & $1.01^{+0.41}_{-0.52}$ & $3.18^{+0.46}_{-0.52}$ & $1.93^{+3.45}_{-\text{---}}$ & $50.91^{+\text{---}}_{-18.91}$ & $0.95^{+\text{---}}_{-0.21}$    & $59.1$ 
\\
PKS\;0558-504		& $29.2^{+10.3}_{-8.4}$		& $0.44^{+0.45}_{-\text{---}}$	& $2.37^{+0.22}_{-0.17}$	& $4.22^{+3.13}_{-2.09}$	& $24.86^{+9.94}_{-\text{---}}$	& $0.94^{+\text{---}}_{-0.41}$	& $20.2$ \\
\hline
\end{tabular}
\end{table*}

Tables~\ref{tab:bplGRhard} and \ref{tab:bplGRsoft} list our best-fitting BPLGRE model results for the $5-7$ keV and $0.5-1$ keV PSDs, respectively. Overall, we do not detect significant differences between the best-fit GRE parameter values in the soft and iron-line bands. The best-fitting heights are smaller in the iron-line band PSDs, although they are still consistent, within the errors, with the soft-band best-fit values. The best-fitting parameter values of the BPL component listed in Tables~\ref{tab:bplGRhard} and \ref{tab:bplGRsoft} are also consistent, within the errors, with the values that we get from the BPL model fit to the observed PSDs (Tables~\ref{tab:bplHard}, \ref{tab:bplSoft}).

The right-hand panels in Fig.~\ref{fig:BestFitExamples} show the best-fitting BPLGRE models to the soft-band PSDs of 1H\;0707-495 and Mrk\;766 (upper and lower panels, respectively). The residual plots of the BPL (shown in the left panels) and BPLGRE models are very similar for both sources. Perhaps a low-amplitude `wavy' pattern in the BPL residuals below $10^{-3}$ Hz (more prominent in Mrk\;766) is diminished in the BPLGRE residuals but, overall, we do not detect significant differences. This is consistent with the fact that the best-fit $\chi^2$ values of both models are comparable. 

In general, the best-fit BPLGRE $\chi^2$ values are smaller than the BPL $\chi^2$ values, but this is expected because the BPLGRE model has a larger number of free parameters. In fact, the best-fitting BPLGRE null hypothesis probability in the case of the iron-line power spectra of NGC\;7314 and Ark\;564 are slightly smaller than the respective BPL best-fitting probabilities (1 and 0.3 per cent, respectively, compared to 1.6 and 0.6 per cent). The same result holds for the soft-band PSDs of NGC\;3516 ($p_{\rm null,BPLGRE}=0.1$ per cent, compared to $p_{\rm null,BPL}=0.3$ per cent). 

The quality of the BPLGRE fits does not change by fixing both the inclination angle to 45$\degr$ and the spin parameter to either one or zero, in all the sources. We repeated the BPLGRE fits to the soft-band PSDs by assuming both spin values, and letting the BH mass of the objects, as well as the X-ray source height, as free parameters. In some objects the best-fit was achieved in the case when $\alpha$ was frozen to zero, and in others when when we kept $\alpha$ frozen to unity (not surprisingly, these were the objects where the best-fit $\alpha$ value was closer to zero and to unity, respectively, as listed in Table~\ref{tab:bplGRsoft}). The resulting best-fitting BH masses are somewhat consistent (within a factor of a few) with the values listed in Table \ref{tab:obj}. However, the resulting best-fit $\chi^2$ values were almost identical to the values listed in Table \ref{tab:bplGRsoft} ($\Delta \chi^2 \approx \pm 1$ in almost all cases). We therefore present below a detailed comparison between the 
goodness of fit of the BPL and BPLGRE models, using the best-fit $\chi^2$ values that are listed in Tables~\ref{tab:bplHard}, \ref{tab:bplSoft}, \ref{tab:bplGRhard} and \ref{tab:bplGRsoft}.

%%%%%%%%%%%%%%%%%%%%%%%%%%%%%%%%%%%%%%%
\section{MODEL FIT COMPARISON}
%%%%%%%%%%%%%%%%%%%%%%%%%%%%%%%%%%%%%%%
\label{sect:modelCompari}

\begin{table*}
 \caption{Comparison of the goodness of fit of the best-fitting models BPL and BPLGRE to the soft band PSDs. The number in parenthesis correspond to the values for the iron-line band.}
 \begin{tabular}{lcccccc}\hline
 &		& 		BPL		&		 BPLGRE	& \\
AGN name		& $F$-test	&  	AIC$_{\rm c,1}$ &  AIC$_{\rm c,2}$  & $\Delta\left[{\rm AIC}_{{\rm c},2}\right]$ & $W\left[{\rm AIC}_{{\rm c},2}\right]$ & $\epsilon_2$\\
		& ratio, prob. (per cent) & &   &   &   \\
\hline
NGC\;7314           &   0.498, 68.6 & 58.324 & 64.813 & 6.489 & 0.038 & 0.039	\\
                    &   (0.255, 85.7)  & (76.591)  & (82.848)  & (6.256) & (0.042) & (0.044)   \\
NGC\;4051           &   0.822, 48.9 & 64.904 & 69.274 & 4.369 & 0.101 & 0.113 	\\
                    &   (1.185, 32.6)  & (56.060)  & (59.733)  & (3.673) & (0.137) & (0.159)   \\
Mrk\;766            &   2.33, 8.7   & 55.452 & 56.186 & 0.734 & 0.409 & 0.693   \\
Ark\;564            &   1.64, 19.65  & 61.317 & 63.205 & 1.888 & 0.280 & 0.389	\\
                    &   (0.378, 76.9)  & (82.090)  & (87.602)  & (5.512) & (0.060) & (0.064)   \\
1H\;0707-495        &   0.910, 44.4 & 75.011 & 78.438 & 3.427 & 0.153 & 0.180 	\\
MCG--6-30-15        &   0.248, 86.3 & 60.682 & 67.244 & 6.562 & 0.036 & 0.038   \\
                    &   (1.660, 18.9)  & (59.634)  & (61.676)  & (2.042) & (0.265) & (0.360)   \\
IRAS\;13224-3809    &   1.520, 22.3 & 49.100 & 52.588 & 3.487 & 0.149 & 0.175	\\
SWIFT\;J2127.4+5654 &   0.510, 67.7 & 74.958 & 80.084 & 5.125 & 0.072 & 0.077	\\
Mrk\;335            &   0.701, 55.6 & 54.703 & 59.959 & 5.256 & 0.067 & 0.072	\\
NGC\;3516           &   0.209, 88.9 & 84.710 & 91.009 & 6.299 & 0.041 & 0.043 	\\
PG\;1211+143        &   0.645, 59.0 & 68.212 & 73.094 & 4.882 & 0.080 & 0.086   \\
PKS\;0558-504       &   1.162, 33.5 & 28.358 & 34.194 & 5.835 & 0.051 & 0.054	\\
\hline
\end{tabular}
\label{tab:bestFitCriteria}
\end{table*}

To decide which model provides the best description of the observed PSDs, the best-fitting results should be compared in a quantitative way. A commonly applied criterion is to convert the $\chi^2$ values obtained for two models into their relative probability measure is the $F$-test. According to \cite{protassov02}, there are two conditions that must be satisfied for the proper use of this test: the two models that are being compared must be nested, and the null values of the additional parameters may not be on the boundary of the set of possible parameter values. In our case, the BPLGRE and BPL models are nested, since the BPL model is a sub-class of the BPLGRE model when the height is very large (strictly speaking, when $h\rightarrow \infty$). At the same time, the `null' values of $h$ are not on the boundary of the possible parameter values (i.e.\ $h=0$, as the height cannot be negative). We therefore use the $F$-test to compare the goodness of fit of the two models. The results (i.e.\ the value of F-
statistic and the corresponding probability) are listed in the second column of Table~\ref{tab:bestFitCriteria}. The null hypothesis in this case is that the more complicated BPLGRE model does not provide a better fit, when compared to the simpler BPL model. We found that $p_{\rm null}>8$ per cent for the model fits to all the iron-line and soft-band PSDs. This result supports the null hypothesis, and indicates that the BPLGRE model does not provide a better fit to the observed PSDs.

We also consider an additional test to investigate which one of the two models describes `best' the observed PSDs. To this end, we calculate the Akaike information criterion, AIC$_{\rm c}$, in its `corrected' version in order to take into account the bias introduced by the finite size of the sample \citep[]{akaike73,sugiura78}, 
\eqb
{\rm AIC}_{\rm{c}}=2k-2C_L +\chi^2 + \frac{2k(k+1)}{N-k-1}
\label{eq:aic}
\eqe
\noindent
$C_L$ is the constant likelihood of the true hypothetical model, and does not depend on either the data or tested models, $k$ is the number of free model parameters, and $N$ is the number of data points in the PSDs. The AIC$_c$ values for both the BPL and BPLGRE best fitting models (AIC$_{\rm c,1}$ and AIC$_{\rm c,2}$, respectively) are listed in Table~\ref{tab:bestFitCriteria} (they are estimated using equation~\ref{eq:aic} without $C_L$, which should have the same value for both models). In general, the model with the lowest AIC$_{\rm c}$ is the `most' preferred model among all models fitted to a given data set. The BPLGRE values, AIC$_{\rm c,2}$, are always larger than the BPL values, AIC$_{\rm c,1}$, indicating that the former model is `less' preferred. 

We then computed the difference, 
\eqb
\Delta\left[{\rm AIC}_{\rm c,2}\right]={\rm AIC}_{\rm c,2}-{\rm AIC}_{\rm c,1}
\label{eq:aicDiff}
\eqe
(in this way, we effectively cancel out the constant term $C_L$). As a general rule, a $\Delta\left[{\rm AIC}_{\rm c,2}\right]$ value smaller than 2 suggests `substantial evidence' for the BPLGRE model (in the sense that both models fit the data at least equally well), $\Delta\left[{\rm AIC}_{\rm c,2}\right]$ values between 3 and 7 indicate that the BPLGRE model has considerably less support, whereas if $\Delta\left[{\rm AIC}_{\rm c,2}\right]>10$ then the BPLGRE model is highly unlikely \citep{burnham02}. The $\Delta\left[{\rm AIC}_{\rm c,2}\right]$ are between 3 and 7 in ten out of the twelve soft-band PSDs, and in three out of the four iron-line band PSDs. This result indicates that the GRE addition to the BPL model is not supported/needed by the observed PSDs. 

To assign a quantitative measure to the statements above, we also computed the `Akaike weight' for the BPLGRE model as follows,
\eqb
{\rm W\left[AIC\right]}_{{\rm c},2}=\frac{{\rm e}^{-\frac{\Delta\left[{\rm AIC}_{{\rm c},2}\right]}{2}}}{{\rm e}^{-\frac{\Delta\left[{\rm AIC}_{{\rm c},1}\right]}{2}}+{\rm e}^{-\frac{\Delta\left[{\rm AIC}_{{\rm c},2}\right]}{2}}}.
\eqe
The weight, ${\rm W\left[AIC\right]}_{{\rm c},2}$, provides a measure of the `strength of evidence' for the BPLGRE model. It effectively gives the probability that the BPLGRE model is the best model between the two models that we used to fit the PSDs. The results are listed in the sixth column of Table~\ref{tab:bestFitCriteria}. They indicate that, in most cases, most of the Akaike weight lies in the BPL model. The BPLGRE weights are less or equal than 0.1 in 8 out of the 12 soft-band PSDs, and in 2 out of the 4 iron-line PSDs. These numbers indicate that, given the observed PSDs, the BPLGRE model has a smaller than 10 per cent chance of being the best one among the two candidate models.

As a final attempt to compare the extent that one model is better than another one, we also estimated the `evidence ratio', for the model with the largest of AIC difference (i.e.\ BLPGRE). This ratio is defined as:  $\epsilon=\rm{W\left[AIC\right]}_{{\rm c},2}/\rm{W\left[AIC\right]}_{{\rm c},1}$, which in our case reduces to   
\eqb
\epsilon_2={\rm e}^{-\frac{\Delta\left[\rm{AIC}_{{\rm c},2}\right]}{2}}.
\eqe
The evidence ratio is a measure of the relative likelihood of BLPGRE versus the BPL model. The evidence ratio values for the BPLGRE model are also listed in the last column of Table~\ref{tab:bestFitCriteria}. In some case, these values are less than 0.1, suggesting that given the observed PSDs, the BPL model is 10 times more likely than BPLGRE. This is not a large value, however, the fact that it holds for the fits we did to many objects in the sample, indicates again that, overall, the BPL model provides a better fit to the observed PSDs (when the extra complexity of the BPLGRE model is taken into account).

%%%%%%%%%%%%%%%%%%%%%%%%%%%%%%%%%%%%%%%
\section{DISCUSSION AND CONCLUSIONS}
%%%%%%%%%%%%%%%%%%%%%%%%%%%%%%%%%%%%%%%
\label{sect:disc}
We studied the observed PSDs of twelve, X-ray bright AGN, in the $0.5-1$ and $5-7$ keV bands, using archival \xmm data. The total net exposure of the EPIC-pn light curves is larger than 350 ks in all cases (and exceeds 1 Ms in the case of 1H\;0707-497). Our main aim was to search for GRE features imprinted in the observed PSDs, due to X-ray illumination of the inner disc, as suggested by P16. We fitted the $5-7$ keV band PSDs of 4 sources, and the $0.5-1$ keV band PSDs of all objects, with a bending power law model and with a BPL model modified by the disc transfer functions of P16. We did not find significant indications of X-ray GR reflection echo signatures in the observed power spectra of the sources in our sample. 

P16 studied the PSD features in the case of X-ray reflection in the $5-7$ keV band assuming a neutral disc, solar iron abundance, and the lamp-post geometry. Strictly speaking, the P16 results are applicable to the $5-7$ keV band PSDs only. However, P16 argued that their predictions should be rather insensitive of the energy band, and of the assumed X-ray/disc geometry, as long as there is a significant reflection component, and the disc response function is characterized by a sharp rise, a plateau, followed by a decline at longer time scales. In this case the corresponding transfer function, $\Gamma(\nu;h,\alpha,\theta)$ (and hence the observed PSD as well), should show a prominent dip and an oscillatory behaviour, with decreasing amplitude at high-frequencies, similar to the disc transfer functions in the lamp-post geometry. The dip amplitude should depend mainly on the ratio of the reflected component over the total flux (i.e.\ continuum plus reflection) in each energy band. If this fraction is 
significant in the soft band, then GR echoes should be present in the observed soft-band PSDs as well.

The fact that the BPLGRE model does not improve the fit in the iron and soft-band PSDs may argue against the P16 assumptions, i.e.\ the X-ray source may not be point-like but has a finite shape, or the disc is ionized, or the iron abundance is not solar. However, the non-detection of the predicted GREs may simply be due to the fact that a BPL model may fit well the observed PSDs in the frequency range where the observed PSDs are best determined (i.e.\ from $10^{-4}$ to $10^{-3}$ Hz), even if they have a BPLGRE shape. The expected best-fit BPL residuals will have a small amplitude, and it may not be be possible to detect them in each source, individually (see discussion in Section~\ref{ssect:objective2}). We therefore estimated the average best-fitting BPL residuals of many sources. In this way we can increase the signal-to-noise ratio and detect low amplitude, but significant, residuals that cannot be detected in the residuals of the individual sources.

%%%%%%%%%%%%%%%%%%%%%%%%%%% FIGURE 4
\begin{figure}
\includegraphics[width=3.2in]{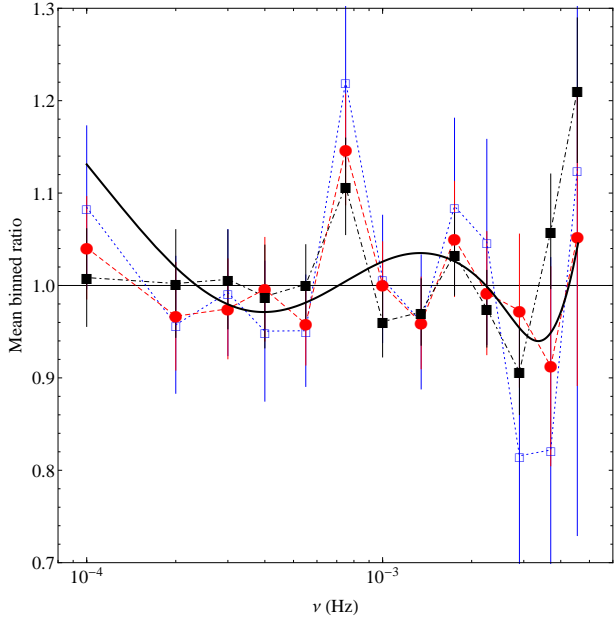}
\caption{Average best-fitting BPL residuals. Open blue squares and red circles indicate the mean residuals for the iron-line band and for the soft band PSDs of six sources, respectively. The filled black boxes indicate the mean binned residuals for the same six sources in the full-band ($0.3-10$ keV) (see Sect.~\ref{sect:disc}). The error bars correspond to the standard deviation of the mean in each bin. The black solid line indicates the theoretical ratio line, in the `highly relativistic' case (see the bottom plot in the lower left panel in Fig.~\ref{fig:models}).}
\label{fig:aveRatios}
\end{figure}

We calculated the best-fitting BPL residuals by dividing the iron-line and the soft-band PSDs by the corresponding best-fitting BPL model. If the BPL model was fitted well to the observed PSDs, the residuals should be consistent with unity at all frequencies. The mean iron-line residuals are plotted in Fig.~\ref{fig:aveRatios}. They are binned in frequency bins which are logarithmically spaced by a factor of $\lambda_{\rm bin}=1.25$ (the residuals are almost identical, irrespective of $\lambda_{\rm bin}$). The open red squares in the same figure indicate the mean soft-band BPL best-model residuals (estimated in the same way) of the sources: NGC\;4051, MCG--6-30-15, Mrk\;766, Ark\;564, 1H\;0707-495, and IRAS\;13224-3809. The BH mass estimates of these sources differ by a factor of less than 3. If there are GRE features in the AGN PSDs, the BPL best-fitting residuals should be different for sources with different BH mass. The differences are such that the residuals do not match even we shift them by a factor 
equal to the ratio of the BH mass of the sources (see for example the differences between the expected BPL best-fitting residuals to the PSDs of a $1.73\times10^{6}$ and $1.73\times10^{7}$ \ms objects in the bottom plots of the left-hand panels in Fig.~\ref{fig:models}). It is for this reason that we decided to estimate the mean residuals only of these six sources only\footnote{The BH estimates for the four sources with iron-band PSDs have similar BH mass estimates. We decided not to include NGC\;7314 during the estimation the mean soft-band residuals, due to its poor quality of its soft-band PSD \citep[see][]{emmanoulopoulos16}.}. 

To increase even further the signal-to-noise ratio, we also considered the $0.3-10$ keV band PSDs (full band PSDs) of the same six sources that we considered in the soft band PSDs. If the ratio of the reflection over the total flux in the full band is similar to the one in the soft and iron-line bands (which are traditionally thought to be reflection dominated), then it would be desirable to also search for the GRE features in the full band PSDs, as in this case the Poisson noise is minimum and we can estimate more accurately the PSDs at high-frequencies. We estimated the full band PSDs as explained in Section \ref{sect:PSDestim}, we fitted them with the BPL model, and we computed the averaged residuals. They are plotted with filled black squares in Fig.~\ref{fig:aveRatios}. Their errors reach a level of $3-4$ per cent at all frequencies going up $2\times10^{-3}$ Hz. Despite the small errors, the average residuals in all bands are consistent with unity: $\chi^2{\rm iron-line}=12.9, \chi^{2}_{\rm soft-band}
=10.2$, and $\chi^2{\rm full-band}=19.3$, for 12 degrees of freedom. This result indicates that the BPL model is fitted well to the observed PSDs and that we cannot detect any significant signature of GRE effects, or any other, deviations. The solid line in Fig.~\ref{fig:aveRatios}  indicates the expected BPL best-fitting residuals in the `highly relativistic' case when $M_{\rm BH}=1.73\times 10^{6}$ \ms\ (as plotted in the bottom plot of the lower left-hand panel in Fig.~\ref{fig:models}). The mean residuals are consistent with this line but, as we have mentioned before, this result is not significant, in the sense that, statistically speaking, the residuals are also consistent with unity. 

We conclude that the X-ray reflection features cannot be detected neither in the soft nor in the iron-line band PSDs of the individual sources. The objects in our sample are among those AGN with the longest X-ray observations with \xmmp. There are no features (with an amplitude larger than $3-5$ percent) that cannot be accounted for by the simple BPL model in the average best-fitting residuals of the brightest, and most variable sources in our sample, even when we consider the full band light curves in order to increase to the maximum possible the signal-to-noise ratio. 

\begin{figure}
\includegraphics[width=3.2in]{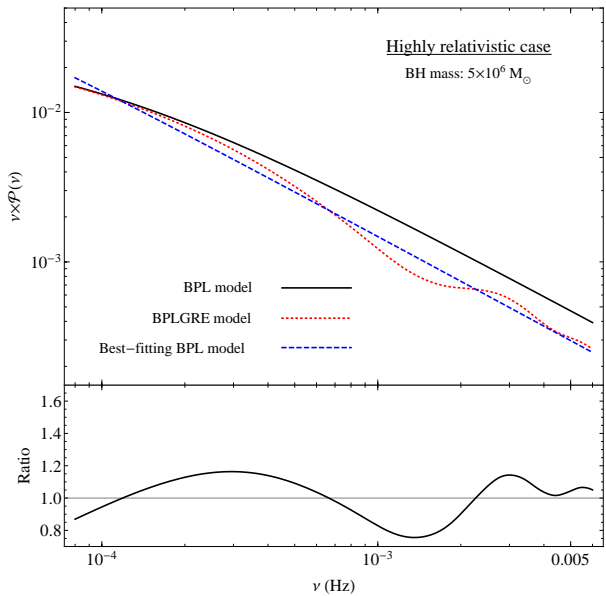}
\caption{The expected GRE features for a BH mass of $5\times10^{6}$ M$_{\odot}$ for the highly relativistic case. The black solid line indicates the BPL model with a high frequency slope of $-2$, and a bend frequency of $8\times 10^{-5}$ Hz. The red dotted line shows the expected `highly relativistic' BPLGRE model PSD (in the case of a Kerr BH and a point-like X-ray source located at a height of $h=3.6$ \rg). The blue dashed lines indicate the best-fitting BPL model to the BPLGRE model PSD, when the high frequency slope is {\it fixed} to $-2$. The attached plot at the bottom shows the ratio of the BPLGRE model PSD over the respective best-fitting BPL model (red dotted and blue dashed lines in the upper panel, respectively).}
\label{fig:model2}
\end{figure}

The main reason for the non-detection of the GRE signatures in the observed PSDs may be lack of sensitivity, i.e.\ the observed PSDs, even when averaged, still have uncertainties that dilute the features needed to be detected in order to claim the corresponding GR effects. However, the biggest uncertainty at the moment is due to the fact that we do not know the exact shape and amplitude of the intrinsic X-ray PSDs in AGN. To demonstrate this issue, we repeat the same procedure as we did in Sect.~\ref{ssect:objective2}). The solid line in Fig.~\ref{fig:model2} shows a BPL model with $s_{\rm h}=2$ and $\nu_{\rm b}=8\times 10^{-5}$ Hz (this would be the bend frequency for a BH mass of $5\times10^{6}$ \ms\ if $\nu_{\rm b}$ would scale inversely with BH mass). The red dotted line indicates the respective `highly relativistic' BPLGRE model (i.e.\ the same BPL model multiplied by the square of the transfer function in the `highly relativistic' case, like the left-hand panels in Fig.~\ref{fig:models}). The blue 
dashed line, in the same figure, indicates the resulting best-fitting BPL model to the theoretical BPLGRE model. However, in this case, the high-frequency slope of the BPL model is kept frozen to the value of $-2$ during the fit. This would be the case if we knew {\it a priori} that this is the intrinsic PSD slope. The bottom panel in Fig.~\ref{fig:model2} shows the ratio of the BPLGRE PSD over the best-fitting BPL model. Clearly, the best-fitting residuals have a much higher amplitude in this case, being of the order of 20 per cent over almost the entire frequency range. As we mentioned above, the amplitude of the average residuals (when we consider all data together) is of the order of $3-5$ per cent. Therefore, we would easily be able to spot the expected GRE features if we knew in advance the intrinsic PSD slope. More work is necessary to investigate the intrinsic PSDs, and then re-examine the agreement of the observed residuals to the expected ones.

%%%%%%%%%%%%%%%%%%%%%%%%%%% FIGURE 5

\section*{ACKNOWLEDGEMENTS}
This work was supported by the `AGNQUEST' project, which is implemented under the `Aristeia II' action of the `Education and Lifelong Learning' operational programme of the GSRT, Greece. It was also supported in part by the grant PIRSES-GA-2012-31578 `EuroCal'. DE and IMM acknowledge the Science and Technology Facilities Council (STFC) for support under grant ST/G003084/1. The research leading to these results has received partial funding from the European Union Seventh Framework Programme (FP7/2007-2013) under grant agreement n.\textsuperscript{\underline{o}}\;312789. This research has made use of NASA's Astrophysics Data System Bibliographic Services. Finally, we are grateful to the anonymous referee for the very useful comments and suggestions that helped improved significantly the quality of the manuscript.

%%%%%%%%%%%%%%%%% APPENDICES %%%%%%%%%%%%%%%%%%%%%

\appendix

\section{IMPLICATIONS Of NON-STATIONARITY TO MODEL FIT RESULTS}
\label{app1:nonStation}
\begin{figure}
\includegraphics[width=2.9in]{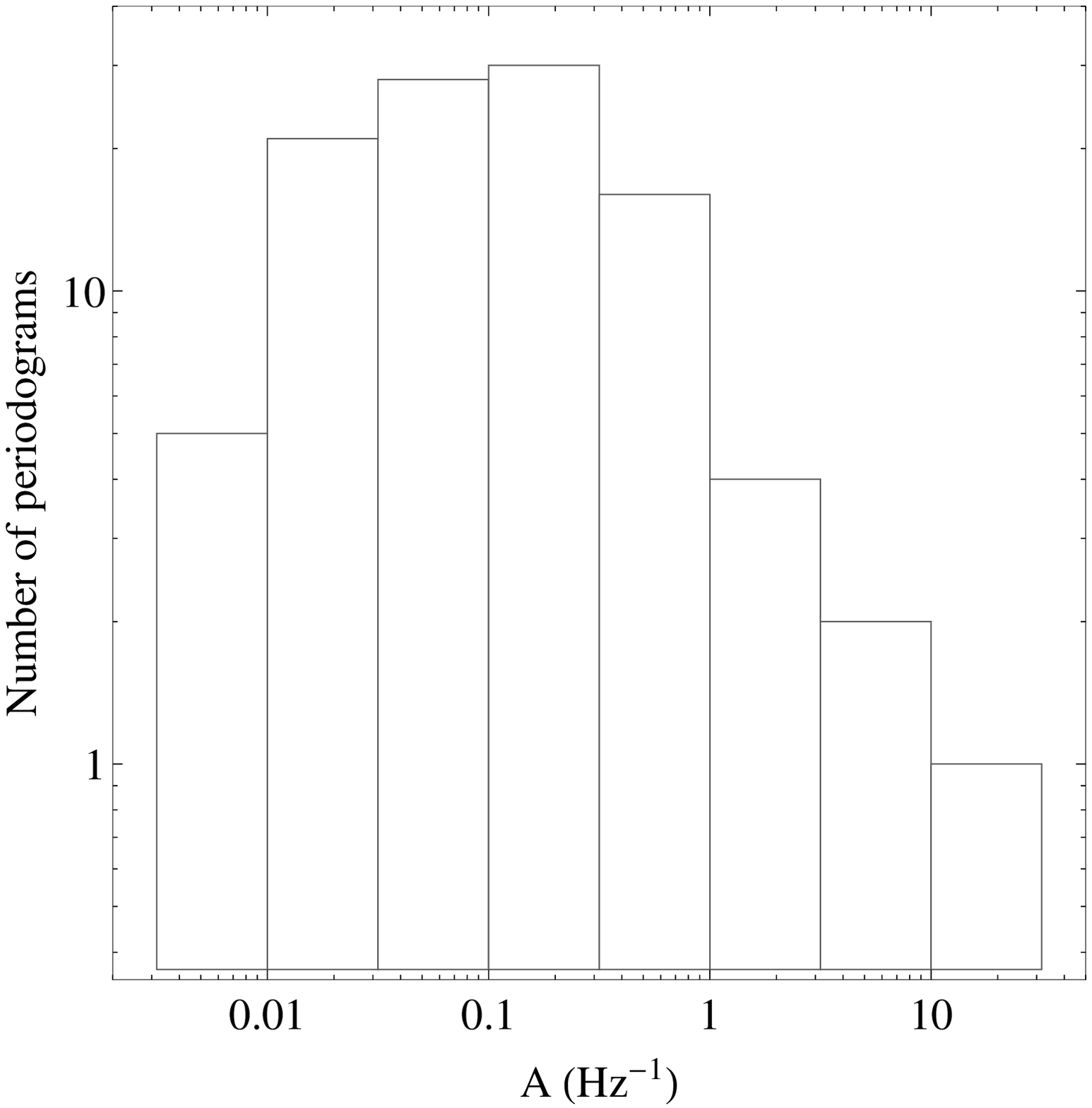}\vspace{0.35em}
\includegraphics[width=2.9in]{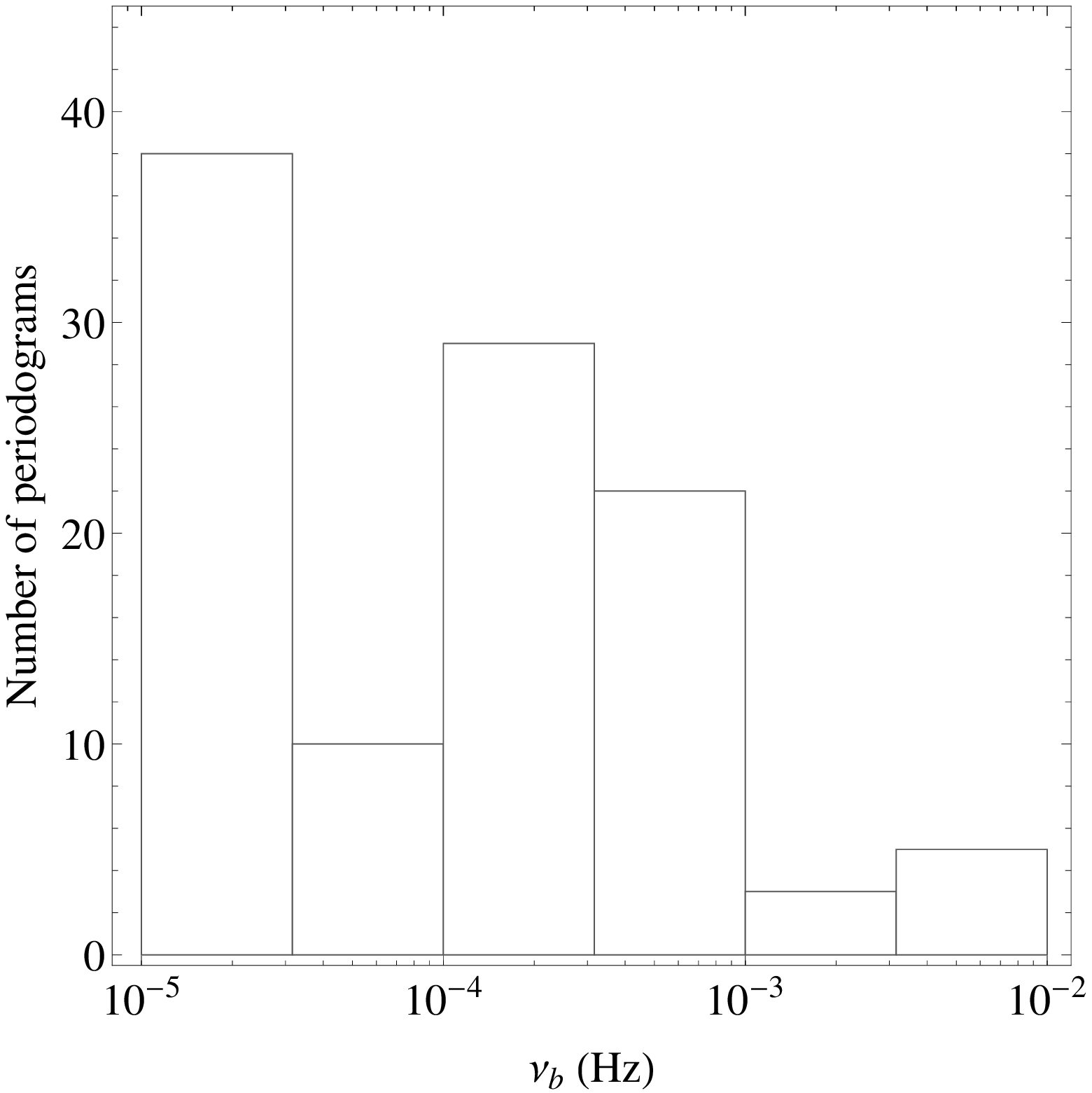}\vspace{0.35em}
\includegraphics[width=2.9in]{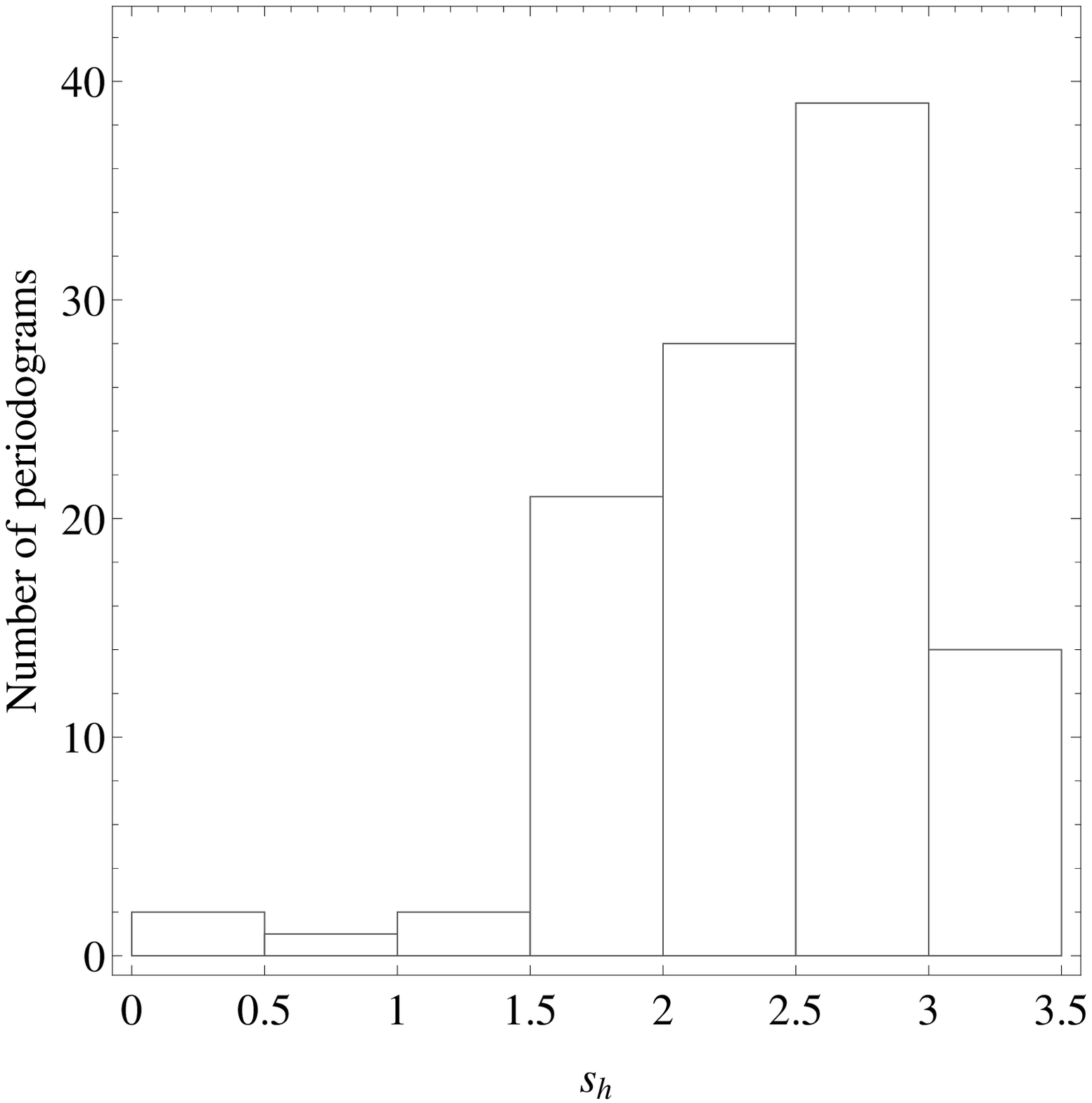}
\caption{The distribution of the BPL model parameter values, from the best-fitting models to all the 121 individual, soft-band  periodograms of 1H\;0707-495 (with Poisson noise), using the maximum-likelihood method.}
\label{fig:maximLikeli}
\end{figure}

In order to investigate the effects on non-stationarity to our best-fitting PSD model results we considered the soft-band PSDs of 1H\;0707-495. This is a highly variable AGN with the longest archival \xmm observations among all sources in the sample. We considered the 121 individual periodograms of this source and we fitted them with a `BPL plus a constant' model (without subtracting the Poisson noise level), using the maximum-likelihood fitting procedure described in appendix~A2 of \citet{emmanoulopoulos13b}. Therefore, we minimize the log-likelihood function $\mathcal{C}=-2\ln \mathscr{L}$, which is equal for the case of even number of data points in the light curves (i.e.\ 100 observations with a bin size of 100 s) to
\eqb
\mathcal{C}&=&2\sum_{j=1}^{j=\frac{N}{2}-1}\left\{\ln\left[\mathscr{P}(f_j;\mathbf{\mathit\gamma})\right]+\frac{P(f_j)}{\mathscr{P}\left(f_j;\mathbf{\mathit\gamma}\right)}\right\}
\nonumber\\[1em]
&&+\ln[{\rm\pi} P(f_{\rm Nyq})\mathscr{P}(f_{\rm Nyq};\mathbf{\mathit\gamma})]+2{\frac{P(f_{\rm Nyq})}{\mathscr{P}\left(f_{\rm Nyq};\mathbf{\mathit\gamma}\right)}.}
\label{eqe:logLikelihood}
\eqe

In Fig.~\ref{fig:maximLikeli} we show the distribution of the BPL best-fitting parameters, i.e.\ $A$, $\nu_{\rm b}$, and $s_{\rm h}$ (top, middle and bottom panels, respectively). The distributions are quite wide but, to a large extend, most of this width must be due to the error in the determination of the individual best-fit values. We did not find the error of each model parameter, for all the 121 model best-fits, because this is a very time consuming procedure, and it is not crucial in our case. 

The sample mean of the $A$, $\nu_{\rm b},$ and $s_{\rm h}$ distributions plotted in Fig.~\ref{fig:maximLikeli} are: $0.26\pm 0.07$ Hz$^{-1}$, $(1.04\pm1.01)\times 10^{-4}$ Hz, and $2.43\pm 0.05$, respectively (the errors correspond to the standard error of the mean). The difference between these values and the best-fitting BPL model parameter values listed in Table \ref{tab:bplSoft} for 1H\;0707-495 is: $ -0.21\pm 0.11$ Hz$^{-1}$, $(1.37\pm 1.12)\times 10^{-4}$ Hz, and $-0.26\pm 0.16$. All three differences are less than 2.3 standard deviations away from zero. 

Consequently, even if the PSD shape is variable with time, and the BPL model parameters vary as widely as the distributions plotted in Fig.~\ref{fig:maximLikeli} suggest (which is certainly not the case in reality), the best-fitting BPL results listed in Table \ref{tab:bplSoft} should correspond to the mean of the intrinsic distribution of the model parameter values. It is in this sense that we claim that fitting the average periodograms can yield the average PSD for a source.  

\section{NON-STATIONARITY AND PSD RESIDUALS}
\label{app2:nonStationPSDresid}
However, even if the best-fitting model parameter values are representative of their mean, if the PSDs are BPL-like, but variable, the average PSD may not have an exact BPL shape. This could explain the large $\chi^2$ values listed in Tables \ref{tab:bplSoft} and \ref{tab:bplHard} for same sources. 

To investigate this issue, we created 10000 simulated light curves, following the method of \citet{emmanoulopoulos13b}. Each light curve has a duration of 100 ks, $\Delta t=100$ s, and a mean which is randomly chosen from the ensemble of the 1H\;0707-495 observed light curve mean values. We assumed an intrinsic PSD which has a BPL shape, and we chose randomly a value from the BPL model parameter distributions shown in Fig.~\ref{fig:maximLikeli} to create each light curve. Consequently, the PSD of each light curve should be different than the others. Then, we created 1000 groups of 10 light curves, chosen randomly from the original sample. We divided them in 10 ks long segments, computed the periodogram for each one, subtracted the Poisson noise level, and calculated the average of the resulting 100 periodograms, exactly as we did for the estimation of the observed PSDs (see Sect.~\ref{sect:PSDestim}). 

\begin{figure}
\includegraphics[width=3.3in]{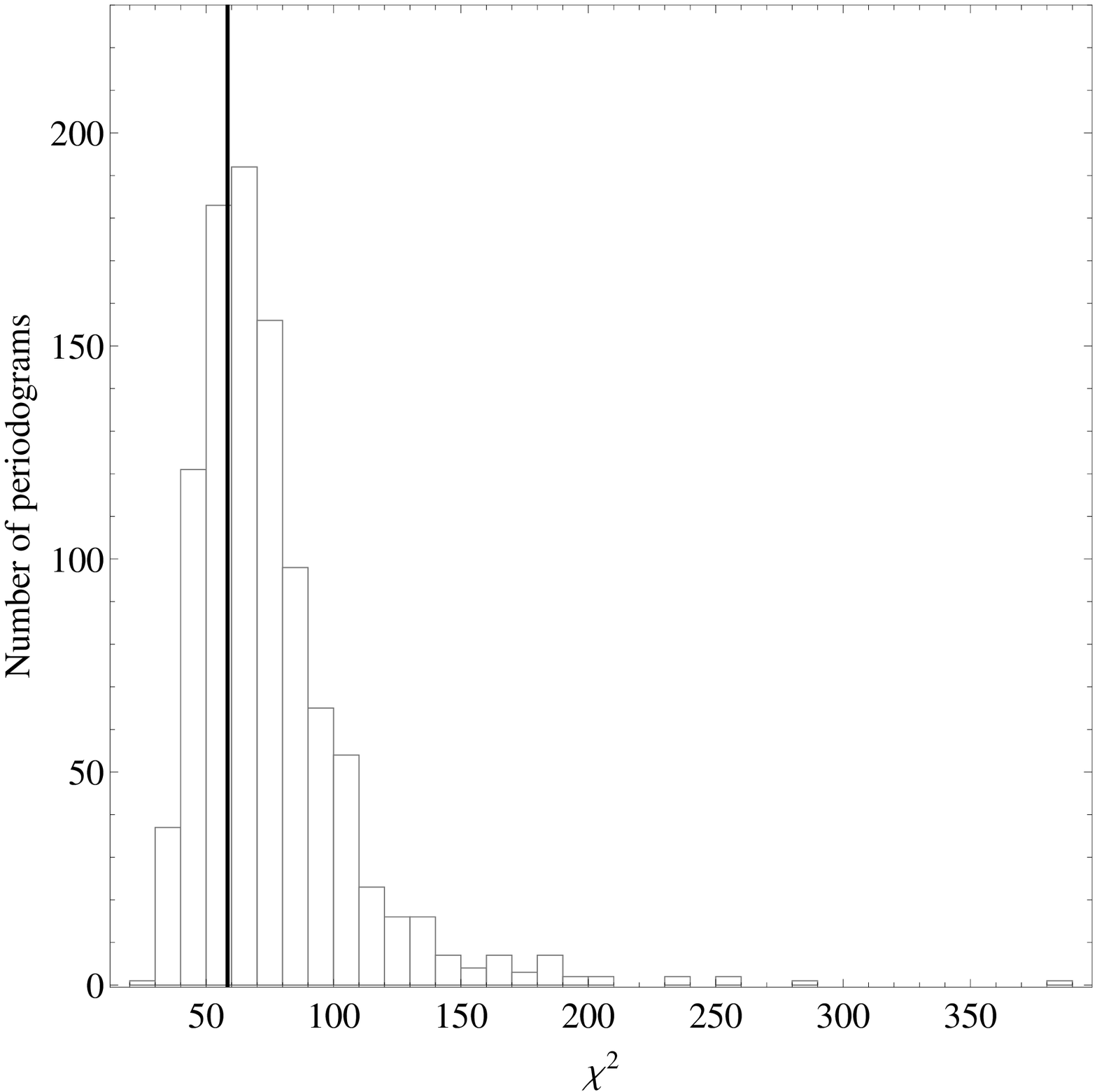}\hspace{1em}
\caption{Distribution of the best-fitting $\chi^2$ values for the best-fitting BPL model to the average periodograms of the simulated light curves (see Sect.~\ref{app2:nonStationPSDresid}). The black line corrsponds to the value of $\chi^2$ equal to 58.4.}
\label{fig:chi2simul}
\end{figure}

We fitted a BPL model to the average periodograms and we recorded the best-fit $\chi^2$. In Fig.~\ref{fig:chi2simul} we show the distribution of these $\chi^2$ values which sometimes seem to be quite large. Clearly, a $\chi^2$ value equal or larger than 68.5 is quite common (this is the best-fitting BPL $\chi^2$ value in the case of the observed, soft-band PSD of 1H\;0707-495; Table~\ref{tab:bplSoft}). The vertical line in Fig.~\ref{fig:chi2simul} indicates this value. According to the distribution of the $\chi^2$ values plotted in Fig.~\ref{fig:chi2simul}, the probability that $\chi^2<58.4$ is {\it less} than 0.3. However, we find that $\chi^2<58.4$ in around 70 per cent of the iron-line and the soft-band BPL best-fits (in 8 out of the 12 sources). Clearly then, the PSDs cannot be as variable as the results plotted in Fig.~\ref{fig:maximLikeli} suggest, for most of the sources in the sample.  On the other hand, non-stationary, variable PSDs may still be possible in the case of SWIFT\;J2127+5654, NGC\;3516, 
and 1H\;0707-495, which show the largest BPL best-fit $\chi^2$ values among the sources (NGC\;7314 and Ark\;564, in the case of the iron-line band PSDs). We plan to investigate this issue further in the future. 

%%%%%%%%%%%%%%%%%%%%%%%%%%%%%%%%%%%%%%%%%%%%%%%%%%
\bsp
\label{lastpage}
\end{document}